\pdfoutput=1

\documentclass[sigconf,nonacm]{acmart}

\setcopyright{none}
\renewcommand\footnotetextcopyrightpermission[1]{}

\usepackage{tikz}
\usepackage{amsmath}

\usepackage{xspace}

\usepackage{filecontents}

\usepackage{multirow}

\usepackage{subcaption}
\usepackage{filecontents}

\usepackage{ifthen}
\usepackage[normalem]{ulem} 
\usepackage{amssymb}

\newboolean{showedits}
\setboolean{showedits}{true} 
\ifthenelse{\boolean{showedits}}
{
	\newcommand{\del}[1]{\textcolor{red}{\sout{#1}}} 
}{
	\newcommand{\del}[1]{} 
	
}

\newboolean{showcomments}
\setboolean{showcomments}{true}
\newcommand{\id}[1]{$-$Id: scgPaper.tex 32478 2010-04-29 09:11:32Z oscar $-$}

\ifthenelse{\boolean{showcomments}}
{\newcommand{\nbc}[3]{
 {\colorbox{#3}{\bfseries\sffamily\scriptsize\textcolor{white}{#1}}}
 {\textcolor{#3}{\sf\small$\blacktriangleright$\textit{#2}$\blacktriangleleft$}}}
 }
{\newcommand{\nbc}[3]{}
 \renewcommand{\del}[1]{} 
 }

\definecolor{ibcolor}{rgb}{0.4,0.6,0.2}
\definecolor{sfcolor}{rgb}{0.2,0.2,0.9}
\definecolor{gncolor}{rgb}{0.8,0.5,0.3}
\definecolor{pwcolor}{rgb}{0.6,0.0,0.6}
\definecolor{cycolor}{rgb}{0.0,0.8,0.2}

\usepackage{wasysym}

\definecolor{todocolor}{rgb}{0.9,0.1,0.1}

\author{Gleb Naumenko}
\email{naumenko.gs@gmail.com}
\affiliation{\institution{University of British Columbia}}

\author{Gregory Maxwell}
\email{greg@xiph.org}
\affiliation{\institution{}}

\author{Pieter Wuille}
\email{pwuille@blockstream.com}
\affiliation{%
  \institution{Blockstream}
}

\author{Alexandra Fedorova}
\email{sasha@ece.ubc.ca}
\affiliation{\institution{University of British Columbia}}

\author{Ivan Beschastnikh}
\email{bestchai@cs.ubc.ca}
\affiliation{\institution{University of British Columbia}}

\begin{document}
\title{Bandwidth-Efficient Transaction Relay in Bitcoin}


\graphicspath{{pictures/}}

\begin{abstract}

Bitcoin is a top-ranked cryptocurrency that has experienced huge
growth and survived numerous attacks. The protocols making up Bitcoin
must therefore accommodate the growth of the network and ensure
security.

Security of the Bitcoin network depends on connectivity between the
nodes. Higher connectivity yields better security. In this paper we make two
observations: (1) current connectivity in the Bitcoin network is too low for
optimal security; (2) at the same time, increasing connectivity will
substantially increase the bandwidth used by the transaction dissemination
protocol, making it prohibitively expensive to operate a Bitcoin
node. Half of the total bandwidth needed to operate a Bitcoin node is currently
used to just announce transactions. Unlike block relay, transaction dissemination
has received little attention in prior work.

We propose a new transaction dissemination protocol, \emph{Erlay}, that not only
reduces the bandwidth consumption by 40\% assuming current connectivity, but
also keeps the bandwidth use almost constant as the connectivity increases. In
contrast, the existing protocol increases the bandwidth consumption linearly
with the number of connections. 
By allowing more connections at a small cost, Erlay improves the
security of the Bitcoin network. And, as we demonstrate, Erlay also
hardens the network against attacks that attempt to learn the origin
node of a transaction.
%
Erlay is currently being investigated by the Bitcoin
community for future use with the Bitcoin protocol.

\end{abstract}

\maketitle

\section{Introduction}
\label{sec:intro}

Bitcoin is a peer-to-peer (P2P) electronic cash
system~\cite{Nakamoto2008Bitcoin}. Recent estimates indicate that there are over
60,000 nodes in the Bitcoin
network~\footnote{\url{https://luke.dashjr.org/programs/bitcoin/files/charts/software.html}}(as of March 2019).
To keep up with the growth in the number of nodes and usage of the network, the
system must be continually optimized while retaining the security guarantees
that its users have come to expect.

Security of the Bitcoin network depends on adequate network
connectivity. Bitcoin literature has repeatedly recommended increasing the
number of connections between nodes to make the network more
robust~\cite{Biryukov2014DeanonBTC, Decker2013BTCPropagation}. As we explain in
Section~\ref{sec:flooding_problematic}, certain attacks become less successful
if the network is highly connected.

Unfortunately, increasing the connectivity of the Bitcoin network linearly
increases the bandwidth consumption of \textit{transaction relay}---the
protocol that currently takes up half of the total bandwidth required to operate
a Bitcoin node. Today, transaction relay alone consumes as much as 18GB
per node per month. If the connectivity were increased from the currently used eight outbound
connections to 24, the per-node bandwidth used for relaying transactions
would exceed 50GB/month. This would make it prohibitively expensive for some users to
operate a Bitcoin node. Despite this inefficiency, transaction relay has not
received much attention in scientific literature, in contrast to block relay~\cite{Corallo2016CompactBlocks,Ozisik2017Graphene,Falcon}.

\begin{figure}[t]
\includegraphics[width=\linewidth]{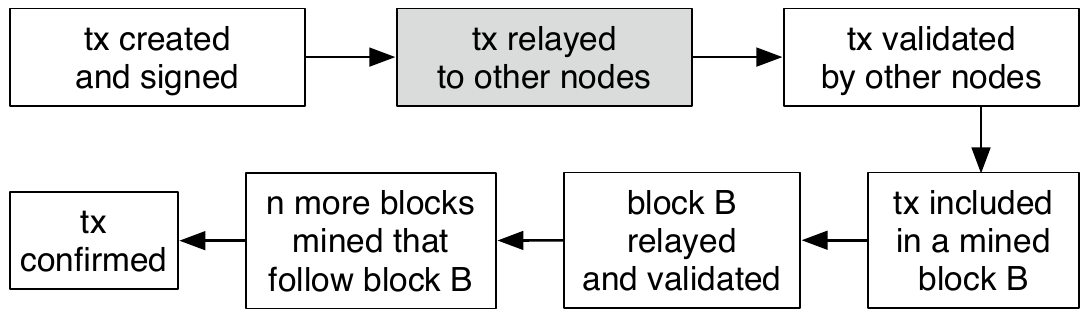}
\caption{Lifecycle of a Bitcoin transaction. In this paper we optimize
  the protocols for relaying transactions between nodes in the Bitcoin
  network (grey box).}
\label{fig:tx_lifecycle}
\end{figure}

The overarching reason why the Bitcoin transaction relay protocol is inefficient
is that it relies on \emph{flooding}.
A Bitcoin \emph{transaction} corresponds to a transfer of funds
between several accounts.
Fig. \ref{fig:tx_lifecycle} overviews the lifecycle of a transaction
in the Bitcoin network. To be accepted by the network of nodes, a
transaction must be first disseminated, or \emph{relayed}, throughout the
network. Then it must be validated and included into a \emph{block}
with other valid transactions. Finally, the block containing the
transaction must be relayed to all the nodes.
Every Bitcoin transaction must reach almost all nodes in the network, and
prior work has demonstrated that full coverage of the network is
important for security~\cite{todd2015sharding}.

Today, Bitcoin disseminates transactions by ensuring that every message received
by a node is transmitted to all of its neighbors. This \emph{flooding} has high
fault-tolerance since no single point of failure will halt relay, and it has low
latency since nodes learn about transactions as fast as
possible~\cite{Spiridoula2013RedunFlood}.

However, flooding has poor bandwidth efficiency: every node in the
network learns about the transaction multiple times.  Our empirical
measurements demonstrate that transaction announcements account for
30--50\% of the overall Bitcoin traffic.  This inefficiency is an
important scalability limitation: the inefficiency increases as the
network becomes more connected, while connectivity of the network is
desirable to the growth and the security of the network.



Prior work has explored two principal approaches to address this inefficient use
of bandwidth. The first is the use of short transaction identifiers (to decrease
message size)~\cite{jl777shortIds2016}. The second is to exclusively use blocks
and never transmit individual transactions~\cite{maxwell2016blocksonly}.
Both approaches are inadequate: short identifiers only reduce the constant
factor and do not scale with the connectivity of the network, while using only
blocks creates spikes in block relay and transaction validation. We discuss
these approaches further in Section~\ref{sec:related}.



\emph{The contribution of this paper is Erlay, a new protocol that
  we designed to optimize Bitcoin's transaction relay while
  maintaining the existing security guarantees.}
The main idea behind our protocol is to reduce the amount of information
propagated via flooding and instead use an efficient set reconciliation method~\cite{Minsky2002PracticalRecon}
for most of the transaction dissemination. In
addition, we design the Erlay protocol to withstand DoS, timing, and other attacks.

We implemented Erlay in a simulator and as part of the mainline
Bitcoin node software, and evaluated Erlay at scale. Our results show
that Erlay makes announcement-related bandwidth negligible while
keeping latency a small fraction of the inter-block interval.
%


In summary, this paper makes the following contributions:

\begin{itemize}

\item We analyze bandwidth inefficiency of Bitcoin's
  transaction relay protocol. We do this by running a node
  connected to the Bitcoin network as well as by running a simulation
  of the Bitcoin network. Our results demonstrate that 88\% of
  the bandwidth used to announce transactions (and around 44\% of the overall
  bandwidth) is redundant.


\item We propose a new, bandwidth-efficient, transaction relay
  protocol for Bitcoin called \emph{Erlay}, which is a combination of 
  fast low-fanout flooding and efficient set reconciliation, designed
  to work under the assumptions of the Bitcoin network.

\item We demonstrate that the protocol achieves a close
  to optimal combination of resource consumption and propagation
  delay, and is robust to attacks. Erlay reduces the bandwidth used
  to announce transactions by 84\% immediately, and allows the Bitcoin network
  to achieve higher connectivity in the future for better security. 
  %

\end{itemize}

Next, we review the background for our work.





%
%



\begin{figure}[t]
\centering
\includegraphics[width=.9\linewidth]{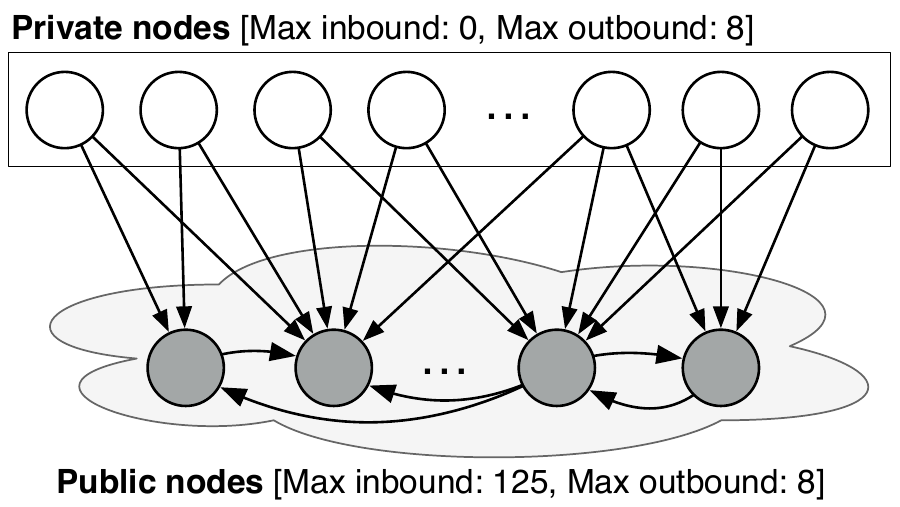}
\caption{Private and public nodes in the Bitcoin network.}
\label{fig:node_types}
\end{figure}

\section{Bitcoin Background}
\label{announcements_explained}

For the purpose of connectivity graph and propagation analysis, there are
2 types of nodes in the Bitcoin network: \textbf{private
nodes} that \emph{do not} accept inbound connections and \textbf{public nodes}
that \emph{do} accept inbound connections (see Fig. \ref{fig:node_types}).
Public nodes act as a backbone of the network: they help new nodes
bootstrap onto the network.
Once they have joined the network, public and private nodes are
indistinguishable in their operation: both node types perform transaction and
block validation, and relay valid transactions and blocks to their
peers.

\begin{figure}[t]
\centering
\includegraphics[width=.8\linewidth]{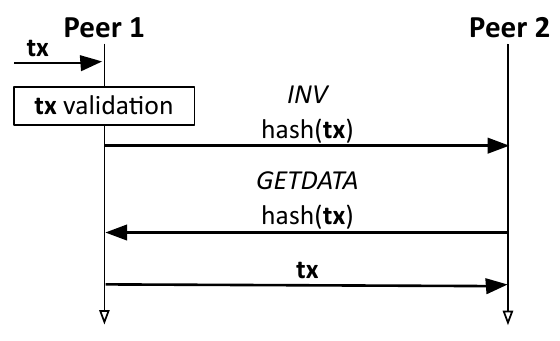}
\caption{Transaction exchange between two peers.}
\label{fig:inv_relay}
\end{figure}

The current version of the Bitcoin transaction relay protocol propagates
messages among nodes using \emph{diffusion}~\cite{core2015commit5400ef}, which
is a variation on random flooding. Flooding is a protocol where each node
announces every transaction it receives to each of its peers. Announcements can
be sent on either inbound and outbound links. With diffusion, a peer injects a
random delay before announcing a received transaction to its peers. This
mitigates timing attacks~\cite{Neudecker2016BTCTimingAnalysis} and significantly
reduces the probability of in-flight collisions
(when two nodes simultaneously announce the same transaction over the link
between them).

The protocol by which a transaction propagates between two peers is illustrated
in Fig. \ref{fig:inv_relay}. When a Bitcoin node receives a transaction (peer 1
in Fig. \ref{fig:inv_relay}), it advertises the transaction to all of its peers
except for the node that sent the transaction in the first place and other nodes
from which it already received an advertisement. To advertise a
transaction, a node sends a hash of the transaction within an \emph{inventory},
or \emph{INV} message. If a node (peer 2 in Fig. \ref{fig:inv_relay}) hears
about a transaction for the first time, it will request the full transaction by
sending a \emph{GETDATA} message to the node that sent it the INV
message.

We refer to the transaction-advertising portion of the protocol (all the INV
messages) as \textit{BTCFlood}. Since it relies on flooding, most transactions
are advertised through each link in the network in one direction (except those
that are advertised during the block relay phase). As a result, a node with $n$
connections will send and receive between $n$ and $2n$ INV messages for a single
transaction (two nodes may announce the same transaction simultaneously to each
other).






Both public and private nodes limit the number of inbound and
outbound connections (Fig. \ref{fig:node_types}). By default a private
node has no inbound connections and up to 8 outbound connections,
while a public node can have 8 outbound connections as well as up to
125 inbound connections (but the inbound connection limit can be configured up to around 1,000).
Thus, as the number of private nodes in the Bitcoin network grows, the
bandwidth and computational requirements to run a public node quickly
increase. This is because private nodes connect to multiple public
nodes to ensure that they are connected to the network through more
than a single peer.

As a result, Bitcoin designers have focused on (1) making the running
of a public node more accessible, in terms of required bandwidth,
computational power, and hardware resources, and (2) making public nodes
more efficient so that they can accept more connections from private
nodes. Our work targets both objectives.


\section{The problem with flooding transactions}
\label{sec:flooding_problematic}

\noindent \textbf{Flooding is inefficient.}
BTCFlood sends many redundant transaction announcements. To see why, let us
first consider how many announcements would be sent if the protocol were
efficient. Since, optimally, each node would receive each announcement exactly
once, \textit{the number of times each announcement is sent should be equal to the
  number of nodes}.

Next, let us consider how many times an announcement is sent with BTCFlood. By
definition, each node relays an announcement on each of the links except the one
where that announcement originally arrived. In other words, each link sees each
announcement once, if no two nodes ever send the same announcement to each other
simultaneously, and more than once if they do. Therefore, \textit{in BTCFlood
  each announcement is sent at least as many times as the number
  of links}.

If $N$ is the number of nodes in the Bitcoin network, the number of links is
$8N$, because each node must make eight outbound connections. Therefore, the
number of \emph{redundant} announcements is at least $8N - N = 7N$. Each announcement
takes 32 bytes out of 300 total bytes needed to relay a single transaction to
one node. (These 300 bytes include the announcement, the response and the full
transaction body). Therefore, if at least seven out of eight announcements are
redundant (corresponding to 224 bytes), at least 43\% of all announcement
traffic is wasteful.


We validated this analysis experimentally. We configured a public Bitcoin node
with eight outbound connections and ran it for one week. During this time, our
node also received four inbound connections. We measured the bandwidth dedicated
to transaction announcements and other transaction dissemination traffic. A
received announcement was considered redundant if it corresponded to an already
known transaction. A sent announcement was considered redundant if it was not
followed by a transaction request. According to our measurements (taken at
multiple nodes at different locations) 10\% of the
traffic corresponding to received announcements and 95\% of the traffic
corresponding to the sent announcements was redundant. Overall, 55\% of all
traffic used by our node was redundant.



\noindent \textbf{Higher connectivity requires more bandwidth.}
Given that the amount of redundant traffic is proportional to the number of
links, increasing the connectivity of the network (the number of outbound links
per node) linearly increases bandwidth consumption in BTCFlood.

\begin{figure}[t]
\centering
\includegraphics[width=.8\linewidth]{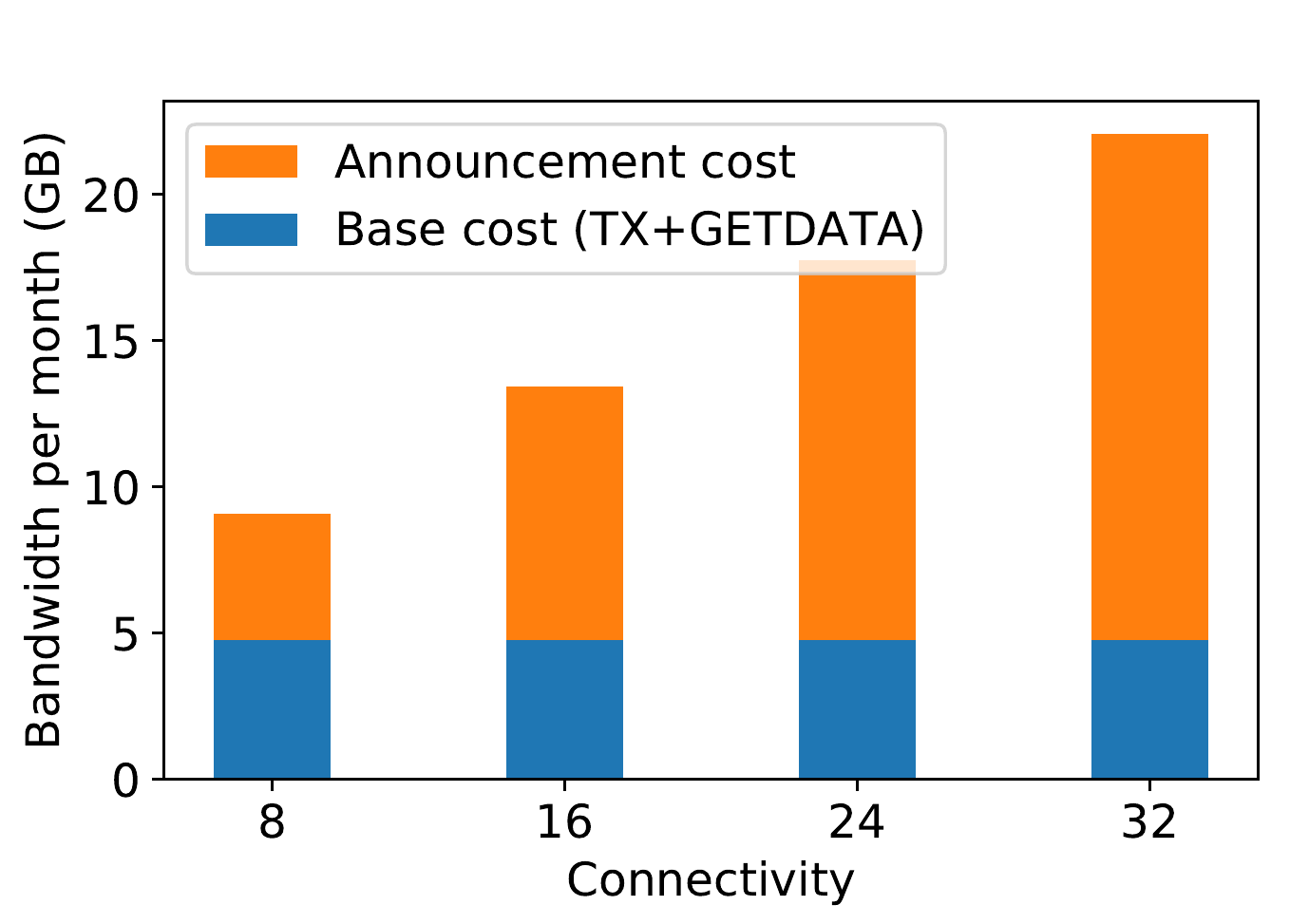}
\caption{Analytical cost of relaying transactions via flooding for one Bitcoin node during one month.
}
\label{fig:scaling_connectivity}
\end{figure}

We modeled how the bandwidth consumption of disseminating one transaction across
the network of 60K nodes increases with connectivity.
Fig.~\ref{fig:scaling_connectivity} (whose results we confirmed via simulation)
shows that announcement traffic turns dominant as the network becomes more
connected. With eight connections per node, a private node may consume 9GB of
bandwidth per month just for announcing transactions.
Setting connectivity to 24 in Bitcoin today would cause 
transaction relay to consume over 15GB/month.



\noindent \textbf{Higher connectivity offers more security.}
In P2P networks, higher connectivity improves network security. This was
demonstrated by both traditional P2P research~\cite{Reka2002NetworksMech,
  Reka2000NetworksTolerance} and Bitcoin-specific prior
work~\cite{Heilman2015Eclipse, Biryukov2014DeanonBTC, Apostolaki2017Hijack,
  Decker2013BTCPropagation, Neudecker2018NetworkAspectsBlockchain}.




Certain attacks become less successful if the network is highly
connected~\cite{Apostolaki2017Hijack, Neudecker2016BTCTimingAnalysis,
  Grundmann2018ExploitingTA}. The eclipse attack paper~\cite{Heilman2015Eclipse} has
shown that fewer than 13 connections would be detrimental to the security of the
network.
A recently discovered vulnerability~\cite{Segura2018Txprobe} relies on
\emph{InvBlock}~\cite{Miller2015DiscoveringB}. InvBlock is a technique that
prevents a transaction from being propagated by first announcing it to a node,
but then withholding the transaction contents for two minutes.
With higher connectivity, this attack is easier to mitigate.

\section{Protocol requirements}
\label{protocol_reqs}


\noindent \textbf{R1: Scale with the number of connections. }
Our main goal is to design a transaction dissemination protocol that 
has good scalability as a function of the number of connections.
This way, we can make the network more secure without sacrificing performance.


\noindent \textbf{R2: Maintain a network topology suited for a decentralized environment.}
Bitcoin's premise of a decentralized environment puts constraints on the design of
its network. Although imposing a structure onto a network, e.g., by organizing
it into a \emph{tree} or \emph{star} topology, or by using DHT-style routing,
enables bandwidth-efficient implementation of flooding, this also introduces the
risks of censorship or partitioning~\cite{Apostolaki2017Hijack}.
The topology of the
network must, therefore, remain unstructured, and routing decisions must be made
independently by every node based on their local state.



\noindent \textbf{R3: Maintain a reasonable latency.}
Transaction propagation delays should remain in the ballpark of those
experienced with the existing protocol. Low latency is essential to user
experience and enables better efficiency in block relay~\cite{Corallo2016CompactBlocks}.


\noindent \textbf{R4: Be robust to attacks under the existing threat model.}
Our protocol must remain robust under the same threat model as that assumed by
the existing protocol. Similarly to Bitcoin, we assume that an attacker has
control over a limited, non-majority, number of nodes in the network, has a
limited ability to make other nodes connect to it, and is otherwise
unrestricted in intercepting and generating traffic for peers that it is
connected to.


The transaction relay protocol must not be any more susceptible to DoS attacks
and client deanonymization, and must not leak any more information about the
network topology~\cite{Neudecker2016BTCTimingAnalysis} than the existing
protocol.

\newcommand{\bobset}{$B$\xspace}
\newcommand{\aliceset}{$A$\xspace}

\begin{figure}[t]
\centering
\includegraphics[width=.7\linewidth]{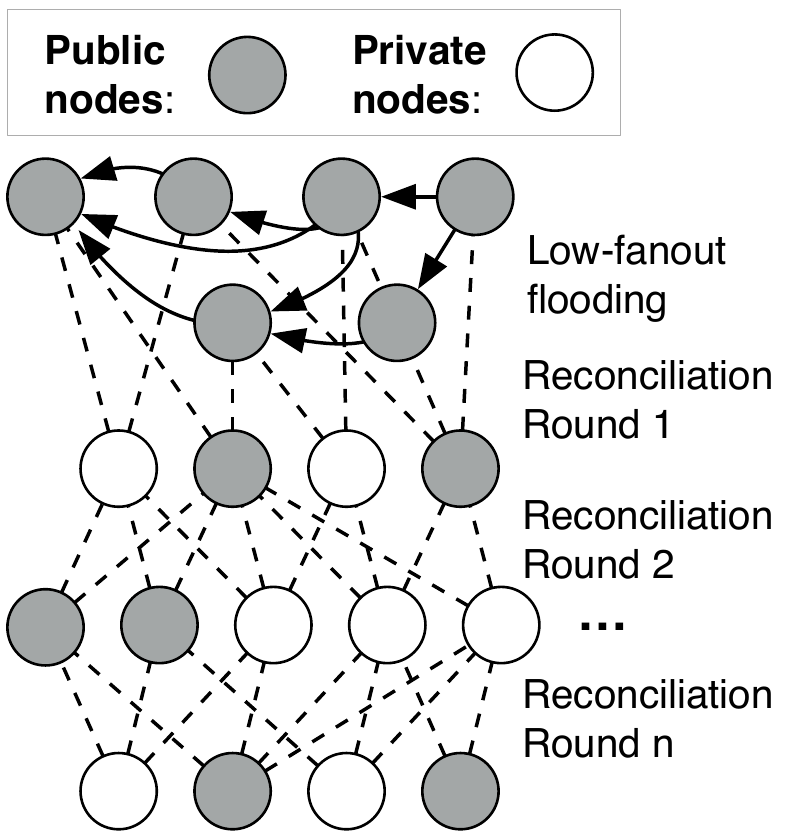}
\caption{Erlay disseminates transactions using low-fanout flooding as
  the first step, and then several rounds of reconciliation to reach
  all nodes in the network.}
\label{fig:erlay_design}
\end{figure}

\section{Erlay design}
\label{sec:design}

Traditionally, P2P networks addressed inefficiency of flooding by imposing a
structured overlay onto an ad-hoc topology. We refrained from structured network
organizations for security reasons discussed in
Section~\ref{protocol_reqs}. Instead, our design relies on two common
system-building techniques: delay and batching.

Instead of announcing every transaction on each link, a node using our protocol
advertises it to a subset of peers---this is called \textit{low-fanout
flooding}. To make sure that all transactions reach the entire network,
nodes periodically engage in an interactive protocol to discover announcements
that were missed, and request missing transactions. This is called \textit{set reconciliation}. Our
protocol, Erlay, is comprised of low-fanout flooding and set reconciliation
(Fig.~\ref{fig:erlay_design}).


\begin{figure}[t]
\centering
\includegraphics[width=.85\linewidth]{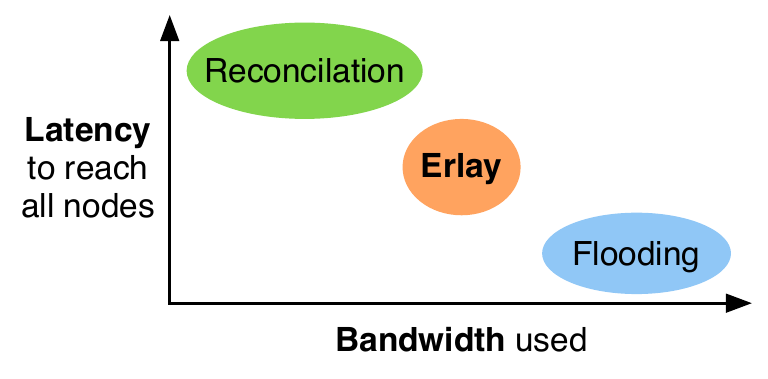}
\caption{Comparison of reconciliation, flooding, and Erlay in their
  bandwidth usage and latency to reach all nodes.}
\label{fig:bw_latency_tradeoff}
\end{figure}

\textbf{Low-fanout flooding.}
The rationale behind low-fanout flooding is to expediently relay a transaction
to be within a small number of hops from every node in the network. If each
transaction ends up close to every node, then reconciliation can finish
dissemination using a small number of rounds. Therefore, a key decision in
low-fanout flooding is to which peers to relay.


%




\textbf{Set reconciliation.}
\emph{Set reconciliation} was proposed as an alternative to
synchronization in distributed
systems~\cite{Minsky2002PracticalRecon}. Using set reconciliation a
node in a P2P network periodically compares its local state to the
state of its peers, and sends/requests only the necessary information
(the state difference). Set reconciliation may be viewed as an
efficient version of \emph{batching} (accumulating multiple state updates
and sending them as a single message). The key challenge in
practical reconciliation is for the peers to efficiently compute their
missing transaction state, and to limit the exchanged transactions to
just those that the other peer is missing.

Fig.~\ref{fig:bw_latency_tradeoff} shows how Erlay attempts to find a sweet spot
in terms of bandwidth and latency by combining flooding, which wastes bandwidth
but disseminates transactions quickly, and reconciliation, which takes longer,
but does not waste bandwidth.


\subsection{Low-fanout flooding}
\label{sec:fanout_design}

Flooding is expensive, so we want to use it sparingly and in \emph{strategic}
locations. For that reason, only well-connected public nodes flood transactions
to other public nodes via outbound connections. Since every private node is
directly connected to several public nodes, this policy ensures that a
transaction is quickly propagated to be within one hop from the majority of the
nodes in the network. As a result, only one or two reconciliation rounds are
needed for full reachability (\textbf{R3}). According to this, the protocol we propose
may be viewed as two-tier optimistic replication~\cite{Saito2005OptimisticReplication}.

To meet our scalability goal (\textbf{R1}), we limit the flooding done by public
nodes to eight outbound connections even if the total number of these
connections is higher. This way, increasing connectivity does not increase
transaction dissemination cost proportionally.

The decision to relay through outbound connections, but not the inbound ones,
was made to defend against timing attacks~\cite{Neudecker2016BTCTimingAnalysis,
  Segura2018Txprobe}. In a timing attack, an attacker connects to a victim and
listens to all transactions that a victim might send on that link (the inbound
connection for the victim). If an attacker learns about a transaction from
multiple nodes (including the victim), the timing of transaction arrival can be
used to guess whether a transaction originated at the victim: if it did then it will
most likely arrive  from the victim earlier than from other nodes. BTCFlood
introduces a diffusion delay to prevent timing attacks. In Erlay, since we do
not forward individual transactions to inbound links, this delay is not
necessary. So this decision favors both \textbf{R3} and \textbf{R4}.

Transactions in the Bitcoin network may originate at both public and private
nodes. In the protocol we propose, nodes do not
relay their transactions via flooding, so the network
learns about the transactions they have originated via reconciliation: private
nodes add their own transactions to the batch of other transactions that
they forward to their peers during reconciliation.
This is used to hide when transactions are originated at private nodes.
If transactions were instead flooded from private nodes, it would be obvious to public
nodes that those transactions must have been created at those nodes, because according
to the chosen flooding policy, this is the only case where a private node floods
a transaction, as they have no inbound links.
Since a private node
forwards its own transactions as part of a batch, as opposed to individually,
a malicious public node is unlikely to discover the origin of a transaction
(\textbf{R4}).

\subsection{Set reconciliation}

In Erlay peers perform set reconciliation by computing a local \emph{set
  sketch}, as defined by the PinSketch algorithm~\cite{Dodis2004Pinsketch}.
A set sketch is a type of set checksum with two important properties:


\begin{itemize}

\item Sketches have a predetermined capacity, and when the number of elements in
  the set does not exceed the capacity, it is always possible to recover the
  entire set from the sketch by \textit{decoding} the sketch.  A sketch of
  \emph{b}-bit elements with capacity \emph{c} can be stored in \emph{bc} bits.

\item A sketch of the symmetric difference between the two sets (i.e.,
  all elements that occur in one but not both input sets), can be
  obtained by XORing the bit representation of sketches of those sets.

\end{itemize}

These properties make sketches appropriate for a bandwidth-efficient set
reconciliation protocol. More specifically, if two parties, Alice and Bob,
each have a set of elements, and they suspect that these sets largely but
not entirely overlap, they can use the following protocol to
have both parties learn all the elements of the two sets:

\begin{itemize}

\item Alice and Bob both locally compute sketches of their sets.

\item Alice sends her sketch to Bob.

\item Bob combines the two sketches, and obtains a sketch of the symmetric difference.

\item Bob tries to recover the elements from the symmetric difference sketch.

\item Bob sends to Alice the elements that she is missing.

\end{itemize}

This procedure will always succeed when the size of the difference (elements
that Alice has but Bob does not have plus elements that Bob has but Alice does
not have) does not exceed the capacity of the sketch that Alice sent.
Otherwise, the procedure is very likely to fail.

A key property of this process is that it works regardless of the actual set sizes:
only the size of the set differences matters.

Decoding the sketch is computationally expensive and is quadratic in the size of
the difference. Because of this, accurately estimating the size of the difference
(Section~\ref{sec:reconciliation_protocol}) and reconciling before the set
difference becomes too large (Section~\ref{sec:reconciliation_schedule}) are
important goals for the protocol.



\subsubsection{Reconciliation round}\label{sec:reconciliation_protocol}
\begin{figure*}[t]
\centering
\includegraphics[width=\linewidth]{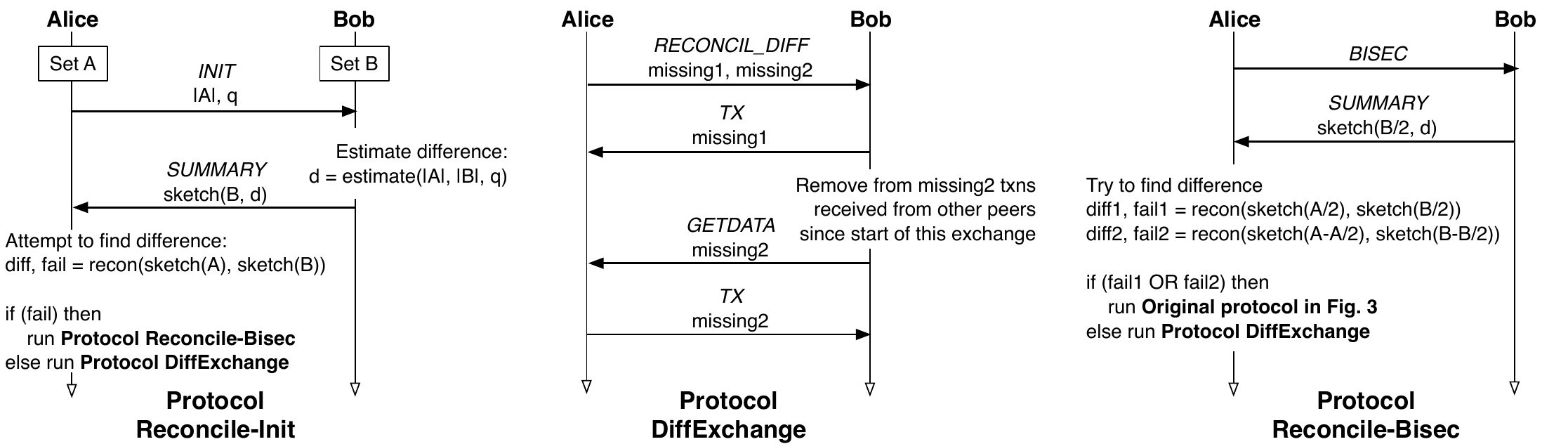}
\caption{Reconciliation protocol with correct difference estimation
  (Reconcile-Init, followed by DiffExchange), and reconciliation protocol
  with incorrect difference estimation (Reconcile-Init, followed by
  Reconcile-Bisec). In case reconciliation fails during Reconcile-Bisec,
  reconciliation falls back to Bitcoin's current exchange method (see
  Fig.~\ref{fig:inv_relay}).
}
\label{fig:recon_scheme}
\end{figure*}

Fig.~\ref{fig:recon_scheme} summarizes the reconciliation protocol.  To execute
a round of reconciliation, every node maintains a \emph{reconciliation set} for
each one of its peers. A reconciliation set consists of short IDs of transactions
that a node would have sent to a corresponding peer in regular BTCFlood, but has
not because Erlay limits flooding. We will refer to Alice's reconciliation set
for Bob as \aliceset and Bob's set for Alice as \bobset. Alice and Bob will
compute the sketches for these reconciliation sets as described in the previous
section.

Important parameters of the protocol are: $D$ -- the true size of the set
difference, $d$ -- an estimate of $D$, and $q$ -- a parameter used to compute
$d$. We provide the derivation of these values below. First, we describe a
reconciliation round:


\begin{enumerate}

  \item According to a chosen reconciliation schedule
    (Section~\ref{sec:reconciliation_schedule}), Alice sends to Bob the size of
    \aliceset and $q$.

  \item Bob computes $d$, an estimate of $D$, between his \bobset and Alice's
    \aliceset (see below).

  \item Bob computes a sketch of \bobset with capacity for $D$
    transactions and sends it to Alice, along with the size of \bobset.

  \item Alice receives Bob's sketch of \bobset, computes a sketch of \aliceset,
    and XORs the two sketches. Now Alice has a sketch of the difference between
    \aliceset and \bobset.

  \item If the difference size was estimated correctly, Alice is able to decode
    the sketch computed in the previous step, request the transactions that she
    is missing from Bob, and then advertise to Bob the transactions that he is
    missing. If the estimation was incorrect (sketch decoding failed), Alice
    will resort to bisection (Section \ref{sec:handling_recon_fail}).

  \item After this process, Alice updates $q$ (see below) and clears
    \aliceset. Bob clears \bobset.

\end{enumerate}

Accurate estimation of $D$ is crucial for success of reconciliation.  Prior work
estimated $D$ using techniques like min-wise hashing~\cite{Broder1997MinWise} or
random projections~\cite{Feigenbaum2003RandomProj}.  These techniques are
complex, and we were concerned that they could end up using more bandwidth than
they save. Therefore, we resorted to a minimalistic approach, where we estimate
the size of the set difference based on just the current sizes of sets and the
difference observed in the previous reconciliation round:

$$d=abs(|A|-|B|) + q \cdot min(|A|,|B|) + c,$$
where $q$ is a floating point coefficient (derived below) that characterizes
previous reconciliation, and $c$ is a parameter for handling special cases.

Indeed, the difference between two sets cannot be smaller than the difference in
their sizes. To avoid costly underestimations, we add the size of the smaller
set normalized by $q$, and a constant $c=1$, which prevents estimating $d=0$
when $|A|=|B|$ and $q \cdot min(|A|,|B|)=0$.

The coefficient $q$ characterizes earlier reconciliation, so before the very
first reconciliation round it is set to zero. At the end of a reconciliation round,
we simply update $q$ based on the true $D$ that we discovered during the round,
by substituting $D$ for $d$ in the above equation, dropping $c$, and then solving for
$q$:

$$q=\frac{D-abs(|A|-|B|)}{min(|A|,|B|)}$$
This updated $q$ will be used in the next reconciliation round. We compute $q$
in this way because we assume that every node in the network will have a
consistent optimal $q$.

Reconciliation is a fertile ground for DoS attacks, because decoding a sketch is
computationally expensive. To prevent these attacks, in our protocol the node
that is interested in reconciliation (and the one that has to decode the sketch)
initiates reconciliation (Alice, in our example). Bob cannot coerce Alice to
perform excessive sketch decoding.


\subsubsection{Reconciliation schedule}
\label{sec:reconciliation_schedule}

Every node initiates reconciliation with one outbound peer every $T$ seconds.
Choosing the right value for $T$ is important for performance and bandwidth
consumption. If $T$ is too low, reconciliation will run too often and will use
more bandwidth than it saves. If $T$ is too high, reconciliation sets will be
large and decoding set differences will be expensive (the computation is
quadratic in the number of differences). A large $T$ also increases the latency
of transaction propagation.



A node reconciles with one peer every $T$ seconds.  Since every node has $c$ outbound
connections, every link in the network would, on average, run reconciliation
every $T \cdot c$ seconds. This means that the average reconciliation set prior to
reconciliation would contain $T \cdot c \cdot {TX}_{rate}$ transactions, where ${TX}_{rate}$
is the global transaction rate. This also means that during the interval between
reconciliations every node would receive
$T \cdot {TX}_{rate}$ transactions. 






We use a value of $1$ second for $T$ in Erlay. With this setting, and the current ratio of
private to public nodes, every public node will perform about eight reconciliations per second.
Given the current maximum Bitcoin network transaction rate ${TX}_{rate}$ of 7 transactions/s, the
average difference set size for this protocol is 7 elements. We evaluate our choice of
parameters in Section~\ref{simulation_results}.

\subsubsection{Bisection for set difference estimation failure}
\label{sec:handling_recon_fail}

Our set reconciliation approach relies on the assumption that an upper
bound for the set difference between two peers is predictable. That
is, if the actual difference is higher than estimated, then
reconciliation will fail. This failure is detectable by a client
computing the difference. An obvious solution to this failure is to
recompute and retransmit the sketch assuming a larger difference in
the sets. However, this would make prior reconciliation transmissions
useless, which is inefficient.





Instead, Erlay uses reconciliation \emph{bisection}, which reuses
previously transmitted information. 
Bisection is based on the assumption that elements are uniformly distributed in
reconciliation sets (this may be achieved by hashing).  If a node is unable to
reconstruct the set difference from a product of two sketches, the node then makes
an additional reconciliation request, similar to the initial one, but this
request is applied to only a fraction of possible messages (e.g., to
transactions in the range $0x0\textrm{--}0x8$). Because of the linearity of sketches, a
sketch of a subset of transactions would allow the node to compute a sketch for
the remainder, which saves bandwidth.

\begin{figure}[t]
\includegraphics[width=\linewidth]{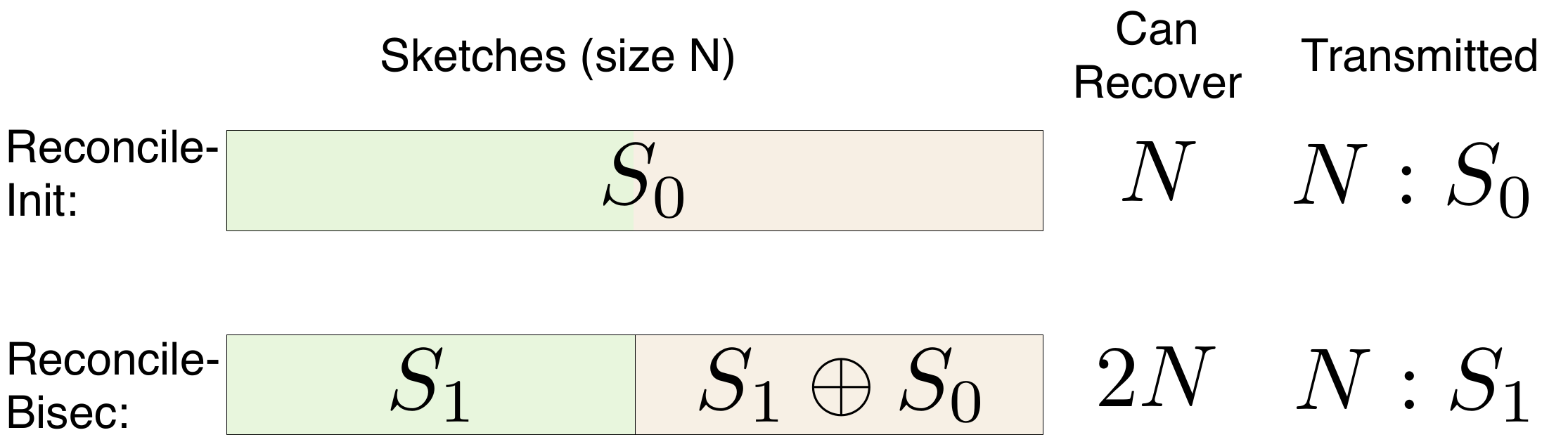}
\caption{Bisection is enabled by the linearity of sketches}
\label{fig:bisection_explained}
\end{figure}

However, this approach would allow recovery of at most $2d$ differences, where
$d$ is the estimated set difference in the initial step.  Even though bisections
are not limited to one and may be applied consequentially without losing
efficiency, in our implementation after a reconciliation step failure we allow
only one bisection with a new overall estimate $2d$ (see
Fig.~\ref{fig:bisection_explained}).
The bisection process is illustrated in protocol Reconcile-Bisec in
Figure~\ref{fig:recon_scheme}.

If bisection fails, then Erlay falls back to the original INV-GETDATA protocol
(Fig.~\ref{fig:inv_relay}) and applies it to all of the transactions in two sets
being reconciled.


\section{Implementation details} \label{recon_impl}


In this section we describe low-level design decisions required to
implement Erlay and increase its bandwidth efficiency (\textbf{R2})
and make it robust to \emph{collision-based} DoS attacks
(\textbf{R4}).

\begin{figure}[t]
\centering
\includegraphics[width=\linewidth]{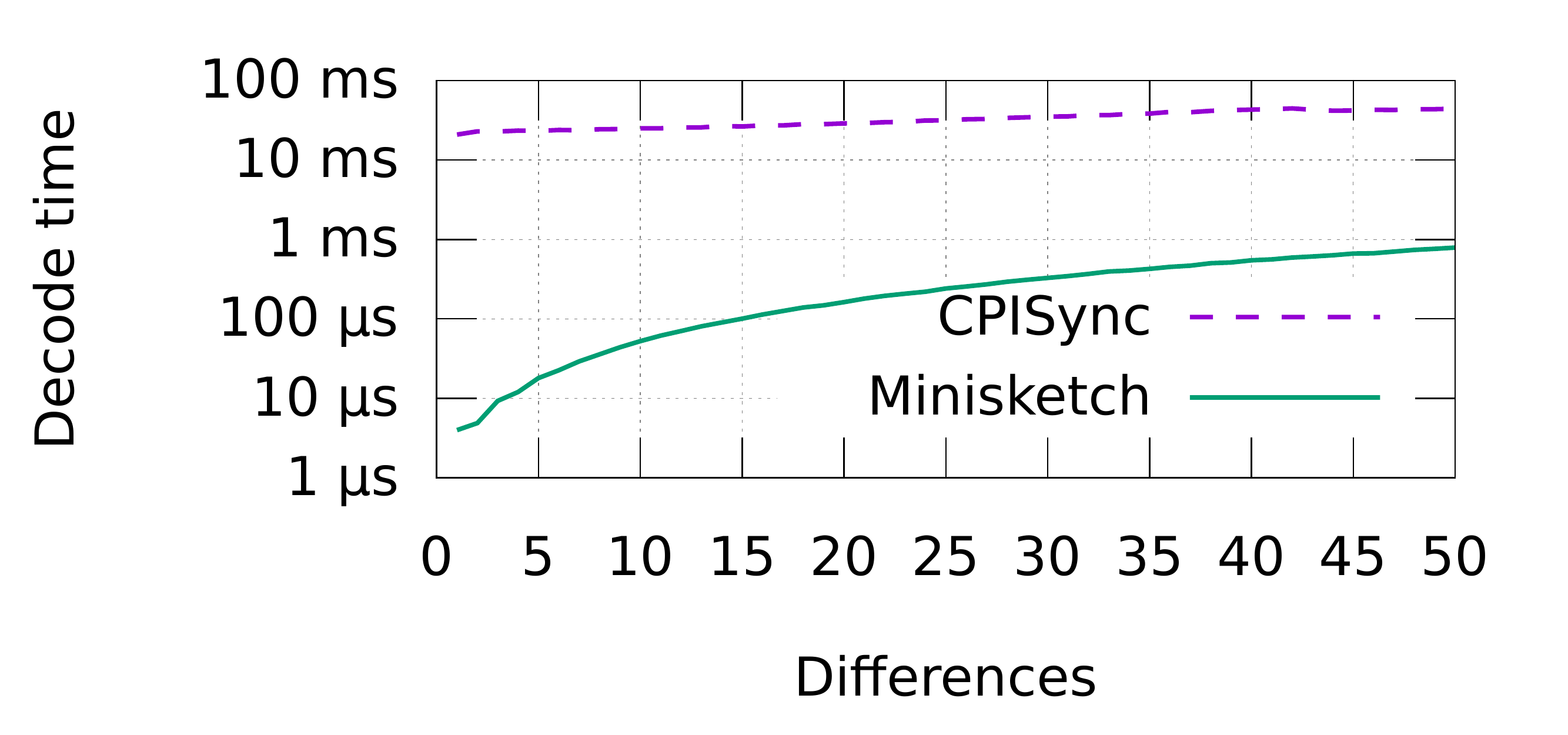}
\caption{The decode time of our library (Minisketch) as compared to
  CPISync for varying set difference sizes.}
\label{fig:libbench}
\end{figure}

\noindent \textbf{Library implementation.}
%
We created Minisketch\footnote{https://github.com/sipa/minisketch},
a C++ library with 3305 LOC, which is an optimized
implementation of the PinSketch~\cite{Dodis2004Pinsketch} algorithm.
We benchmarked the library to verify that set reconciliation would not
create high computational workload on Bitcoin nodes.
Fig.~\ref{fig:libbench} shows the decoding performance on an Intel
Core i7-7820HQ CPU of our library (Minisketch) as compared to
CPISync~\cite{Trachtenberg2002CPISync}\footnote{\url{https://github.com/trachten/cpisync}}
for varying difference sizes.
Our library has sub-millisecond performance for difference sizes of
100 elements or fewer. As we will show later
(Fig.~\ref{fig:diffs_distr}) this performance is sufficiently fast for
the differences we observe in practice (in simulation and in
deployment).

We used this library to build a reference implementation of Erlay as
a part of the Bitcoin Core software, which we evaluate in
Section~\ref{sec:ref_impl_res}.

\noindent \textbf{Short identifiers and salting.} The size of a transaction ID
in the Bitcoin protocol is 32 bytes. To use PinSketch~\cite{Dodis2004Pinsketch}, we have to use
shorter, 64 bit, identifiers.
Using fewer bits reduces the bandwidth usage by 75\% (\textbf{R2}),
but it also creates a probability of collisions. Collisions in transaction relay are an attack surface,
because a malicious actor may flood a network with colluding
transactions and fill \emph{memory pools} of the nodes with
transactions, which would then be propagated and confirmed in a very slow
manner.  Thus we want to secure the protocol against
such attacks (\textbf{R4}).

While collisions on one side of a communication are easy to detect and
handle, collisions involving transactions on both sides may cause a
significant slowdown.  To mitigate this, we use different
salt (random data added to an input of a hash-function) while hashing
transaction IDs into short identifiers.

The salt value is enforced by the peer that initiates the connection,
and per Erlay's design, requests reconciliation. Since the peer
requesting reconciliation also computes the reconciliation difference,
the requestor peer would have to deal with short IDs of unknown
transactions.  Since salt is chosen by the requestor, re-using the same salt
for different reconciliations would allow him to compare salted short IDs
of unknown transactions to the IDs received during flooding from
other peers at the same time.


\noindent \textbf{Low-fanout diffusion delay.}  Bitcoin flooding mitigates
timing attacks~\cite{Neudecker2016BTCTimingAnalysis} and in-flight
collisions by introducing a random delay into transaction
announcements.
For timing attacks Bitcoin assumes that an attacker connects
(possibly, multiple times) to the node (or takes over a fraction of
outbound connections of the node). In a low-fanout model, this attack
is not feasible, because transactions are flooded
through outbound connections only.

In-flight collisions are also not possible in the case of
low-fanout relay through only outbound links, because transactions
are always announced in the same direction of a link.

In consideration of these arguments as well as to reduce latency, Erlay has a lower
random diffusion interval. Instead of using $T_{oi} = 2$ seconds for
outbound connections and $T_{ii} = 5$ seconds for inbound, Erlay uses
$T_{oi} = 1$ seconds for outbound.

\noindent \textbf{Reconciliation diffusion delay.}
Even though in Erlay timing attacks by observing low-fanout flooding
are not feasible, an attacker would be able to perform
them through reconciliations.  To make timing attacks through
reconciliations more expensive to perform, we enforce every peer to
respond to reconciliation requests after a small
random delay (in our implementation, a Poisson-distributed
random variable which is on average $T_{ri} = 1$ seconds),
which is shared across reconciliation requests from all peers,
and we rate-limit reconciliations per peer. This measure would make Erlay
better than BTCFlood at withstanding timing attacks.

Our measure in Erlay has the same idea as in flooding/low-fanout diffusion;
however, having the ratio $T_{ii}/T_{oi}$ higher makes timing
attacks less accurate, because during $T_{ii}$ (the average time
before an attacker receives a transaction) a transaction would be
propagated to more nodes in the network.

We chose the interval of 1 seconds because a lower interval would make Erlay
more susceptible to timing attacks than Bitcoin,
and a higher interval results in a high latency.

\section{Evaluation methodology} \label{eval_method}



In evaluating Erlay we focus on answering the following three
questions:

\begin{enumerate}

\item How does Erlay compare to BTCFlood in latency (the time that it takes for
  the transaction to reach all of the nodes) and bandwidth (the number of
  bits used to disseminate a transaction)?

\item How do the two parts of Erlay (low-fanout flooding and
  reconciliation) perform at scale and with varying connectivity,
  varying number of nodes, and varying transaction rates?

\item How do malicious nodes impact Erlay's performance?

\end{enumerate}

We use measurement results from two sources to answer the questions
above.
First, we used a simulator to simulate Erlay on a single
machine (Section~\ref{simulation_results}).
Second, we implemented Erlay in the mainline Bitcoin client
and deployed a network of Erlay clients on the Azure cloud across
several data centers (Section~\ref{sec:ref_impl_res}).

\textbf{Simulator design.}
Our simulation was done with ns3. We modified an open-source Bitcoin
Simulator~\cite{gervais2016security} to support transaction relay.
The original simulator had 9663 LOC; the version we modified has 9948 LOC.

Our simulator is based on the INV-GETDATA transaction relay protocol
(see Section~\ref{announcements_explained}). It is parameterized by
the current ratio of public nodes to private nodes in the Bitcoin
network and the transaction rate based on the historical data from the
Bitcoin network (7 transactions per second on average).
We simulate the different ratios of faults in the
network by introducing Black Hole nodes, which receive transactions
but do not relay them.

Our simulator does not account for heterogeneous node resources, the block relay
phase, the joining and leaving of nodes during the transaction relay phase
(churn), and does not consider sophisticated malicious nodes.

The propagation latency measured for BTCFlood by our simulator matches
the value suggested for the validation of
Bitcoin simulators~\cite{Fadhil2016MeasSym},
and our measured bandwidth matches our analytical estimates.

\textbf{Topology of the simulated network.}
We emulated a network similar to the current Bitcoin network, since
inferring the Bitcoin network topology is
non-trivial~\cite{Neudecker2016BTCTimingAnalysis}.
In our simulation we bootstrapped the network in two phases: (1) public nodes
connected to each other using a limit of eight outbound connections, then (2)
private nodes connected to eight random public nodes. In some experiments we
increased connectivity, as indicated in the experiment's description.

Unless stated otherwise, our simulation results are for a network of
6,000 public nodes and 54,000 private nodes (this is the scale of
today's network\footnote{\url{https://bitnodes.earn.com/}\\
\url{https://luke.dashjr.org/programs/bitcoin/files/charts/software.html}}).
In each experiment we first
used the above two steps to create the topology, then we relayed
transactions for 600 seconds (on average, we generated 4,200 transactions
from random private nodes).

\section{Simulation results} \label{simulation_results}

In this section we use simulation to demonstrate latency,
bandwidth consumption, and security of Erlay and compare them to
BTCFlood. 

\subsection{Relay bandwidth usage}

\begin{figure}[t]
\centering
\includegraphics[width=\linewidth]{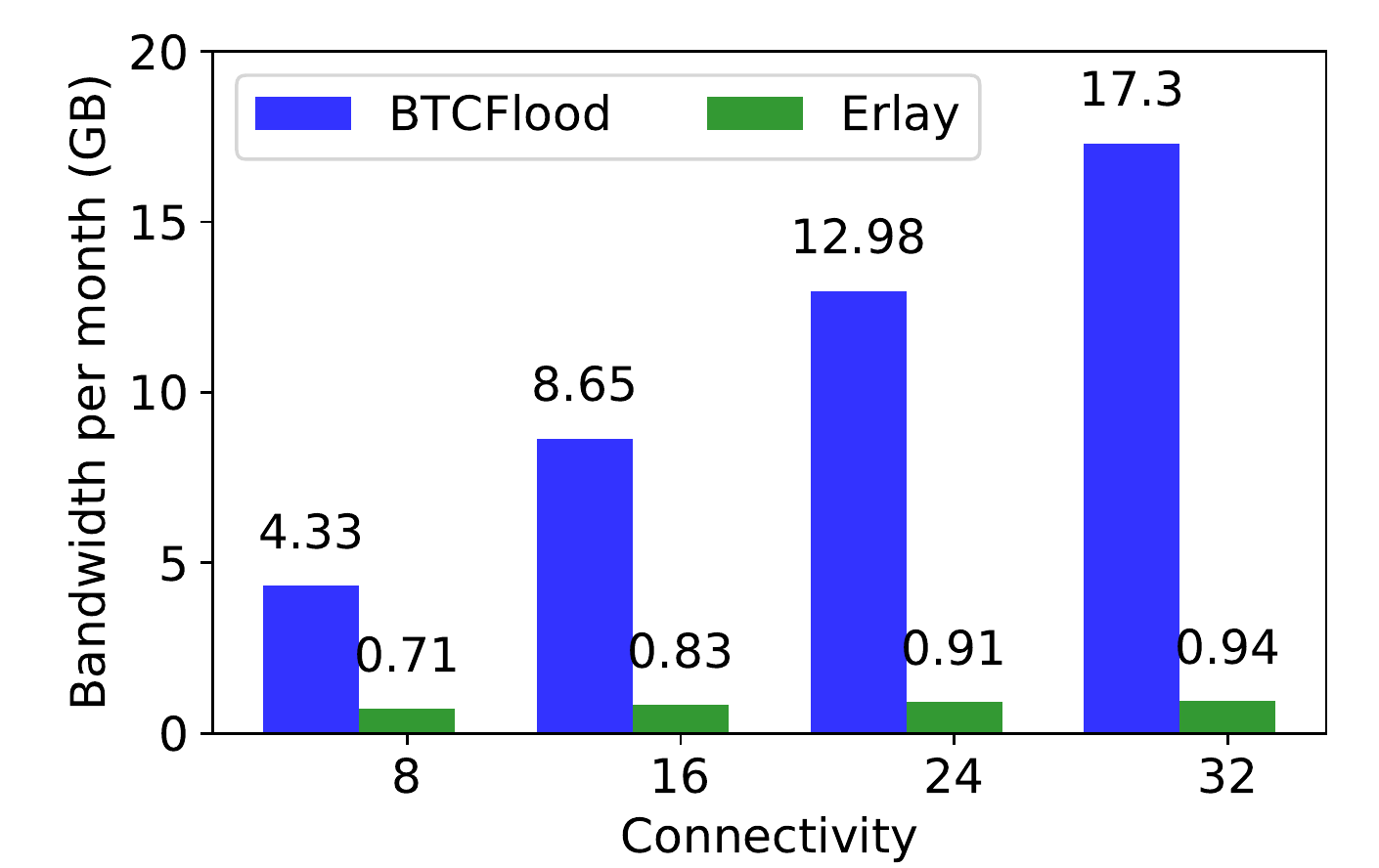}
\caption{Average bandwidth one Bitcoin node spends
  per month to \textbf{announce} transactions.}
\label{fig:bandwidth_connectivity}
\end{figure}

To verify that Erlay scales better than BTCFlood as the connectivity increases,
we varied the number of outbound connections per node and measured the bandwidth
used for announcing transactions. Figure~\ref{fig:bandwidth_connectivity} shows
the results.


With BTCFlood, relay bandwidth increases linearly with the connectivity because
BTCFlood announces transactions on \emph{every} link in the network. With Erlay,
however, bandwidth consumption grows significantly
slower. Erlay seamlessly embraces higher connectivity, which allows for better
security.


\begin{figure}[t]
\centering
\includegraphics[width=\linewidth]{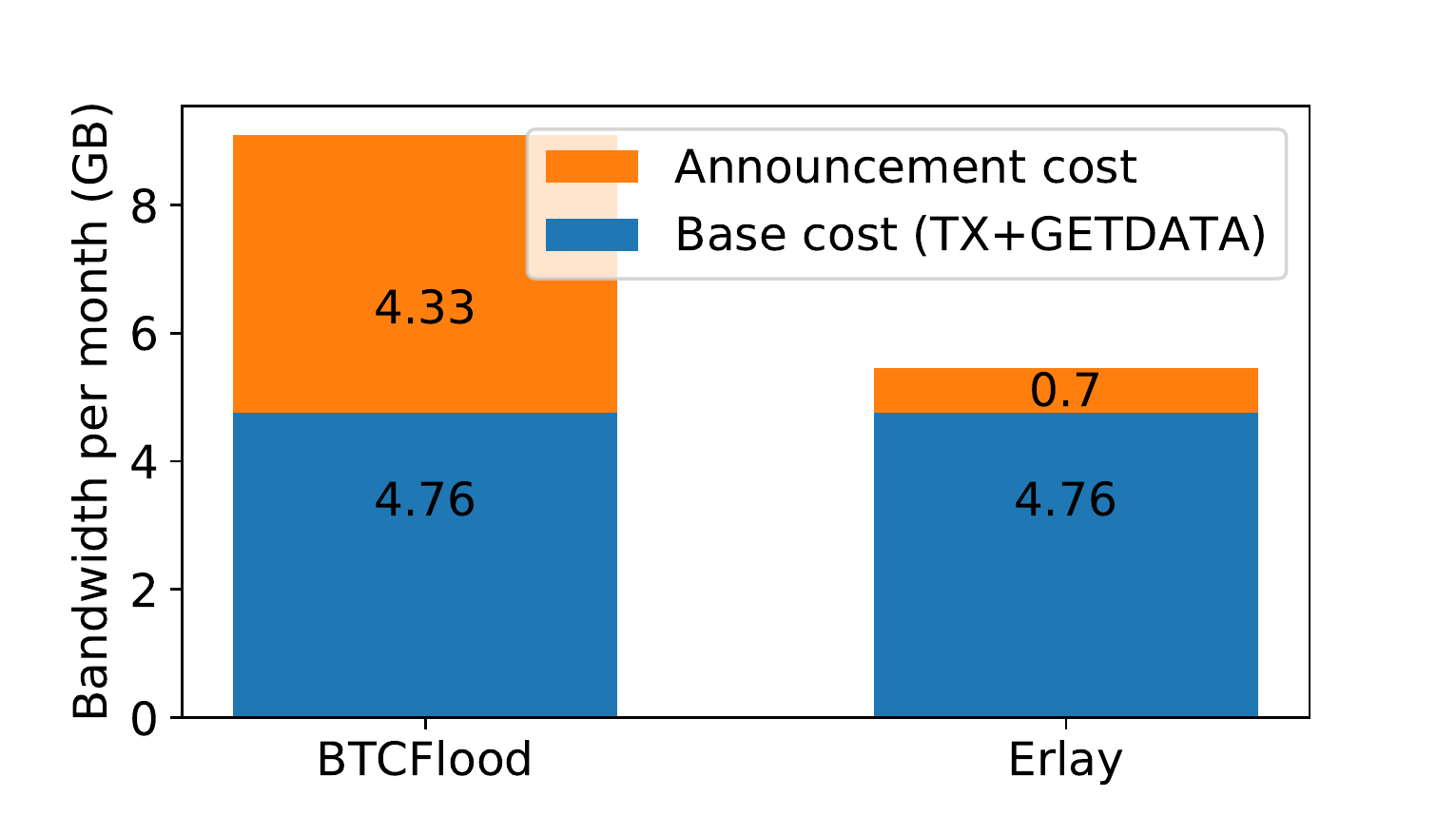}
\caption{Average bandwidth cost of \textbf{fully relaying} transactions during 1 month
  for a Bitcoin node with outbound connectivity of 8.}
\label{fig:scaling60k}
\end{figure}

\textbf{Transaction announcements in overall bandwidth.}
To demonstrate that Erlay's announcement optimization impacts overall
bandwidth, we measure the bandwidth consumed by a simulated network to
relay transactions with BTCFlood and with Erlay.
Fig.~\ref{fig:scaling60k} plots the results for simulations in which
every node establishes 8 connections. Erlay's announcement bandwidth
is just 12.8\% of the relay bandwidth, while for BTCFlood the
announcement bandwidth is 47.6\%.


\textbf{Breaking down Erlay's bandwidth usage.}
To further understand Erlay's bandwidth usage, we broke it down by the
different parts of the protocol: low-fanout flooding, reconciliation, and
post-reconciliation announcements.

Table~\ref{tab:relay_aspects} lists the results. 
%
%
The table shows that about a third of the bandwidth is used by
reconciliation, while low-fanout flooding accounts for a majority of the
bandwidth.
%
%
The post-reconciliation INVs account for a small fraction of Erlay's
bandwidth. 

\begin{table}[t]
\centering
\caption{Breakdown of bandwidth usage in Erlay.} 
\begin{tabular}{r|r}
\textbf{Erlay component}          & \textbf{Bandwidth \%}\\ 
\hline
Low-fanout flooding      & 54\%                          \\ 
\hline
Reconciliation           & 32\%                          \\ 
Bisection                & 0.7\%                          \\ 
Fallback                 & 4.3\%                        \\ 
\hline
Post-reconcile. INVs & 9\%                        \\ 
\hline
\textbf{Total}           & \textbf{100\%}             \\
\end{tabular}
\label{tab:relay_aspects}
\end{table}

\begin{figure}[t]
\centering
\includegraphics[width=.75\linewidth]{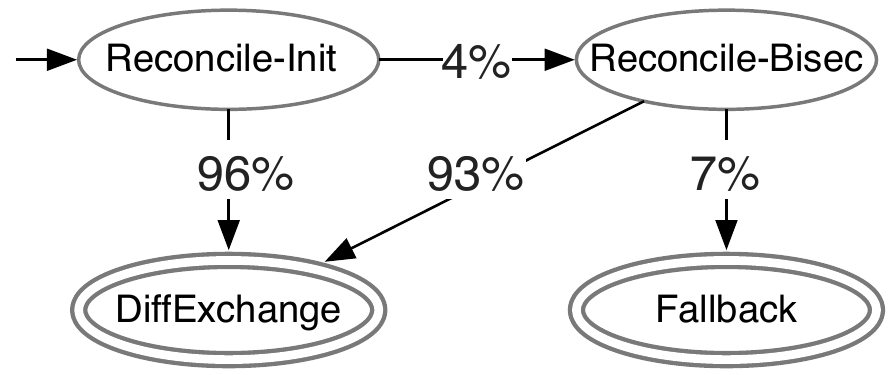}
\caption{Finite state machine of the protocol in
  Fig.~\ref{fig:inv_relay} annotated with transition percentages
  observed in our experiments.}
\label{fig:fsm-protocol}
\end{figure}

\textbf{Set reconciliation effectiveness.} To understand the
effectiveness of Erlay's set reconciliation, we measured how often
reconciliation or the following bisection protocol fail.
Fig.~\ref{fig:fsm-protocol} reports the results aggregated from one of
our simulation runs with 60,000 nodes. The end-to-end probability of
reaching fallback is below 1\%. Since bisection does not introduce
additional bandwidth overhead (while fallback does), the overall
reconciliation overhead is low.

\begin{figure}[t]
\centering
\includegraphics[width=\linewidth]{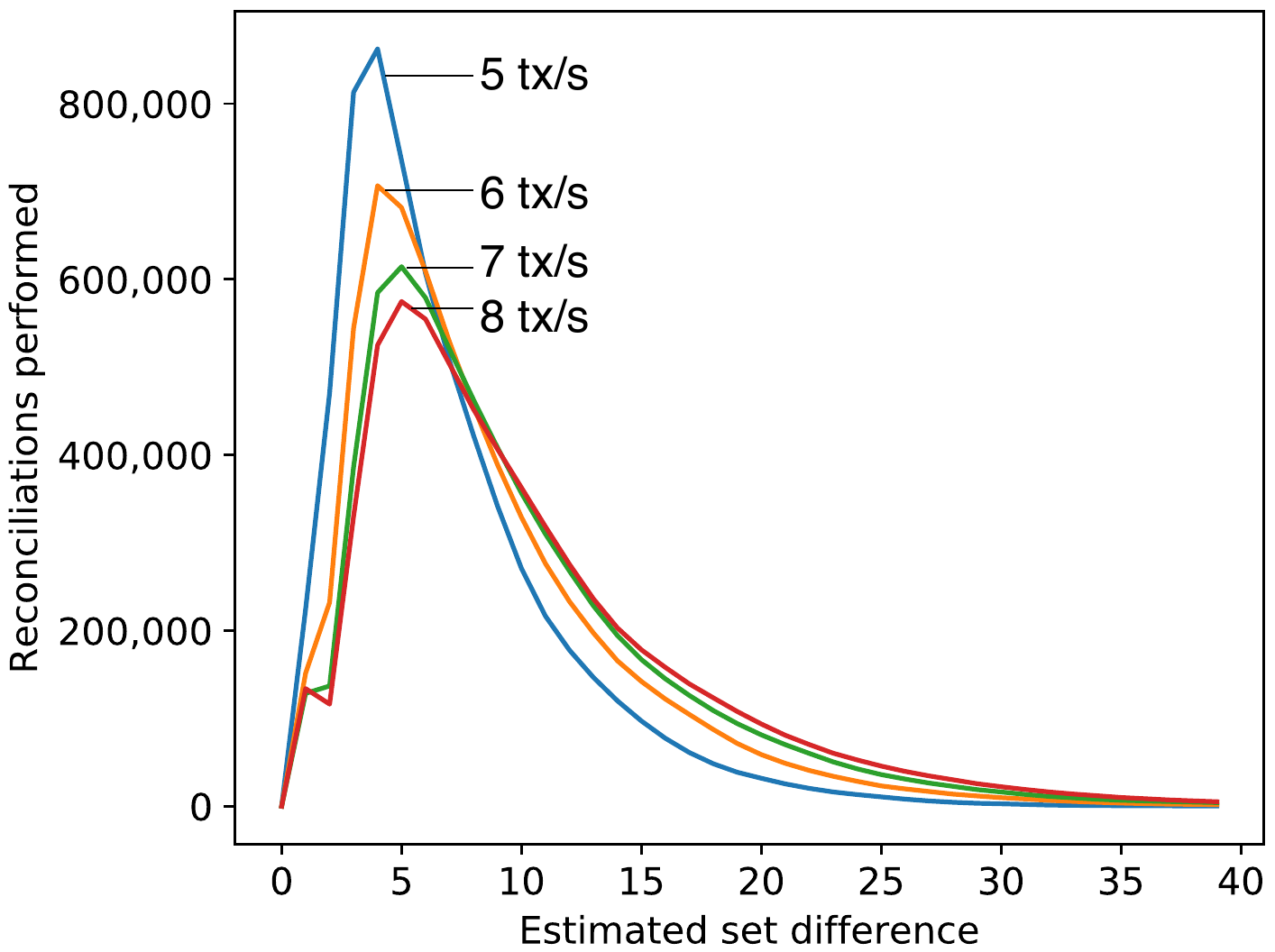}
\caption{Distribution of the set difference estimates during
  reconciliation for different transaction rates.}
\label{fig:diffs_distr}
\end{figure}

Since every reconciliation round requires a set difference estimation,
we measured the distribution of the estimated difference
sizes. Fig.~\ref{fig:diffs_distr} demonstrates that set difference
depends on transaction rate.
This is expected: for the same reconciliation intervals, a higher
transaction rate would result in both reconciling parties receiving
more transactions and would lead to a larger set difference. This
dependency between set difference and transaction rate allows accurate
set difference estimation. Fig.~\ref{fig:fsm-protocol} illustrates
that Erlay's estimate is correct 96\% of the time. For the cases where
Erlay over-estimates and the initial reconciliation fails, the
resulting bandwidth overhead constitutes 9\% of the overall bandwidth.


In our library benchmarks the decode time for a sketch containing 100
differences is under 1 millisecond (Fig.~\ref{fig:libbench}).  Thus,
the computational cost of operating over sketches with the
distribution in Fig.~\ref{fig:diffs_distr} is negligible.




\subsection{Relay latency} 

\begin{figure}[t]
\centering
\includegraphics[width=\linewidth]{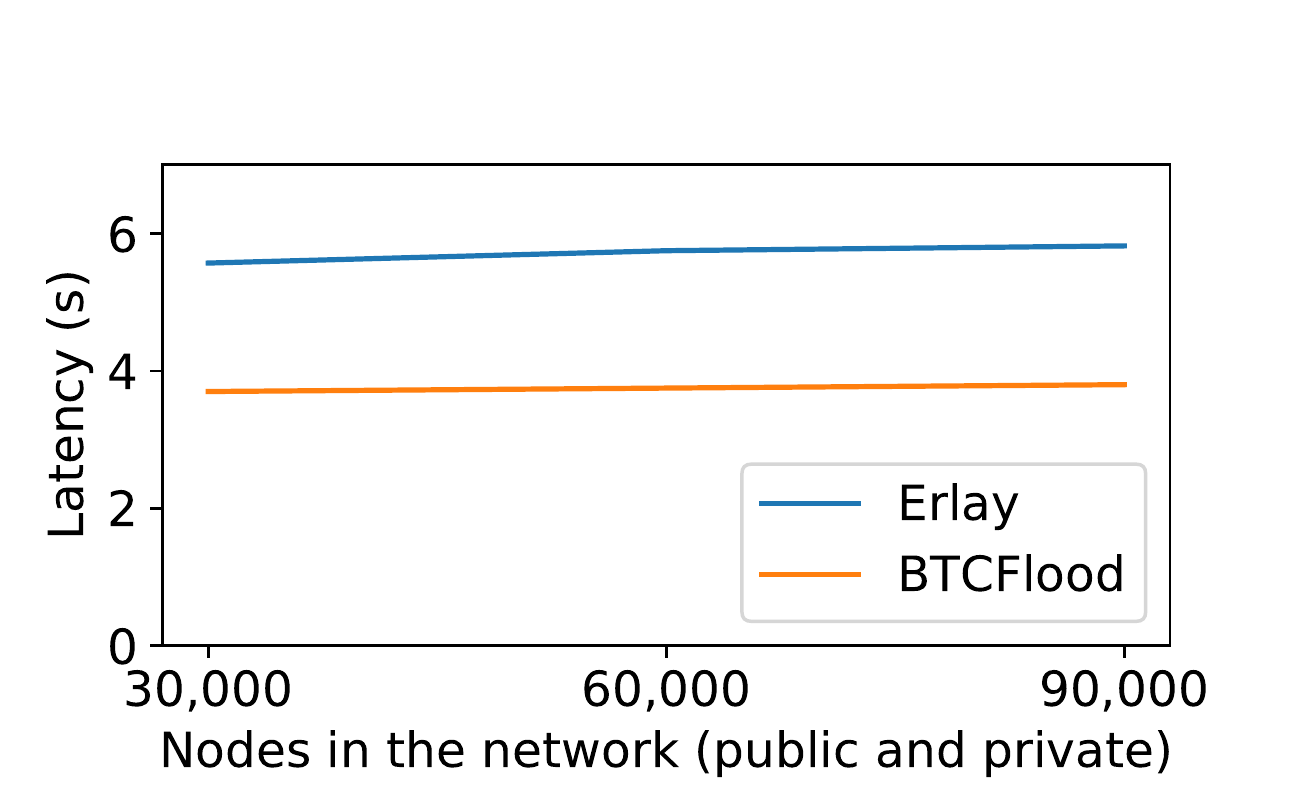}
\caption{Average latency for a single transaction to reach 100\% nodes
  in networks with different sizes.}
\label{fig:latency_scale}
\end{figure}

Fig.~\ref{fig:latency_scale} plots the average latency for a single
transaction to reach all nodes for Erlay and BTCFlood as we vary the
total number of nodes. In this set of experiments we kept constant the
ratio between private and public types of nodes at $9:1$ (this is the
ratio in today's Bitcoin network).
Erlay has a constant latency overhead on top of BTCFlood that is due to its use
of batching. However, this overhead is just 2.6 seconds and changes at
approximately the same rate with the number of nodes as BTCFlood's
latency. Erlay's per-transaction latency can be reduced at the cost of higher
bandwidth usage. This is a tunable parameter, subject to design constraints.

We chose to pay this latency overhead, because this is acceptable
cost to maximize bandwidth efficiency, as we demonstrate in Section~\ref{sec:discussion}.

\begin{figure}[t]
\centering
\includegraphics[width=\linewidth]{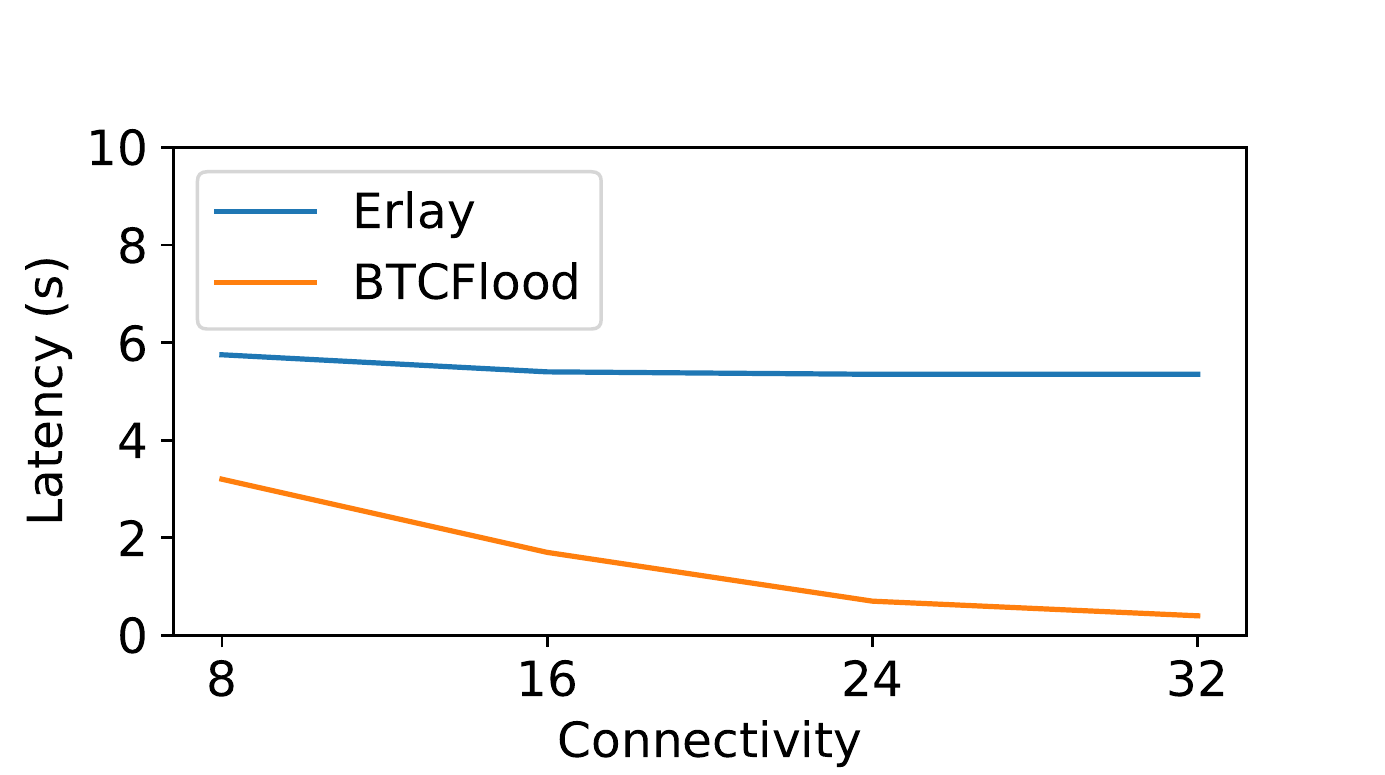}
\caption{Average latency for a single transaction to reach 100\% nodes
  in the network with variable connectivity.}
\label{fig:latency_connectivity}
\end{figure}

One of Erlay's goals is to enable higher connectivity. We therefore
analyzed the latency of Erlay and BTCFlood for different
connectivities of the network. Figure~\ref{fig:latency_connectivity}
demonstrates that, as the connectivity increases,
latency significantly decreases for BTCFlood (at high
bandwidth cost), and only slightly decreases for Erlay without
significant effect on bandwidth.



\begin{figure}[t]
\centering
\includegraphics[width=\linewidth]{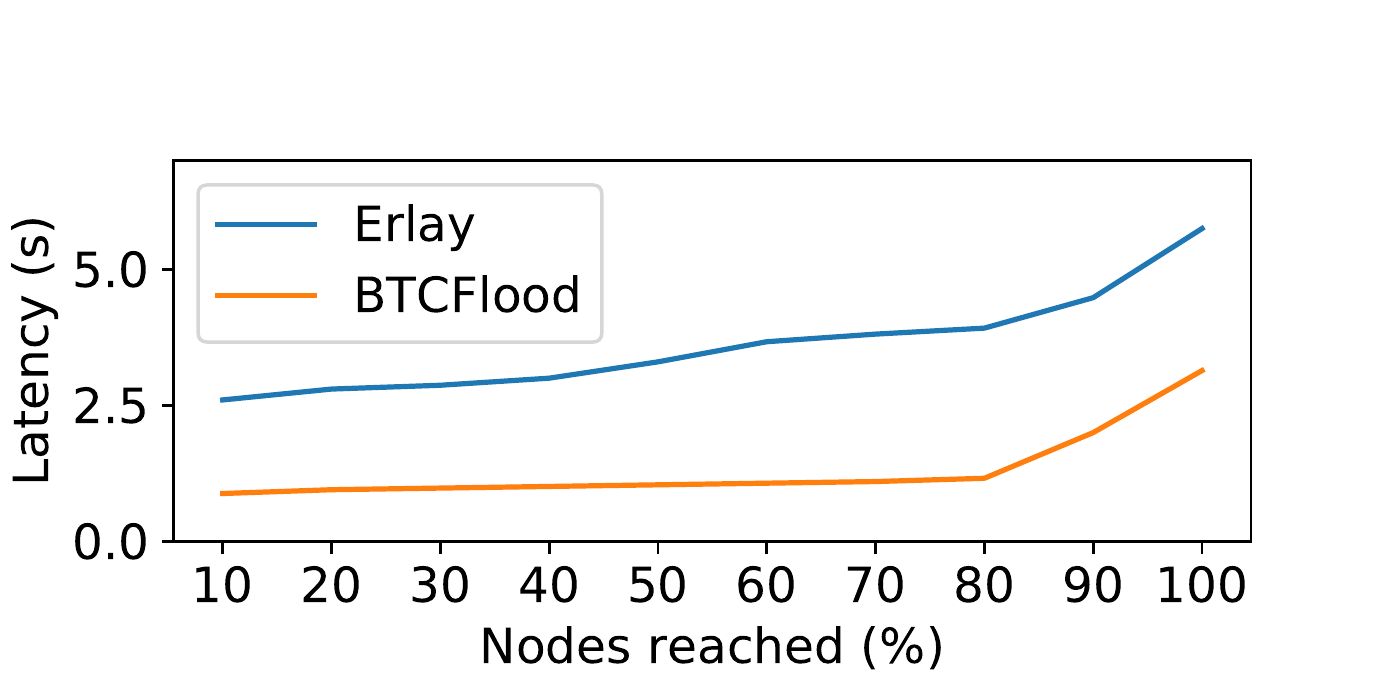}
\caption{Average latency for a single transaction to reach a certain
fraction of nodes in the network}
\label{fig:latency_full}
\end{figure}

To understand how transactions propagate across the network, we
measured the latency to reach a certain fraction of nodes in the
network.  Figure~\ref{fig:latency_full} demonstrates that Erlay
follows the same propagation pattern as BTCFlood with a fairly
constant overhead of 2.6 seconds.

\textbf{Latency under faulty condition}
We also evaluated Erlay's latency in a simple adversarial setting. For
this we simulated a network in which 10\% of the public nodes are
\emph{black holes} and measured the time for a transaction to reach
all nodes. While it is difficult to outperform the robustness of
BTCFlood, an alternative protocol should not be dramatically impacted
by this attack.

According to our measurements, while the slowdown with BTCFlood in
this setting is 2\%, the slowdown with Erlay is 20\%.  We believe that
this latency increase is acceptable for a batching-based protocol. We
have ideas for heuristics that might be applied to mitigate black-hole
attacks and make Erlay less susceptible. For example, a node might
avoid reconciling with those outbound connections that regularly
provide the fewest new transactions.

\subsection{Scalability with transaction rate} \label{sec:withstand_timing}

\begin{figure}[t]
\centering
\includegraphics[width=\linewidth]{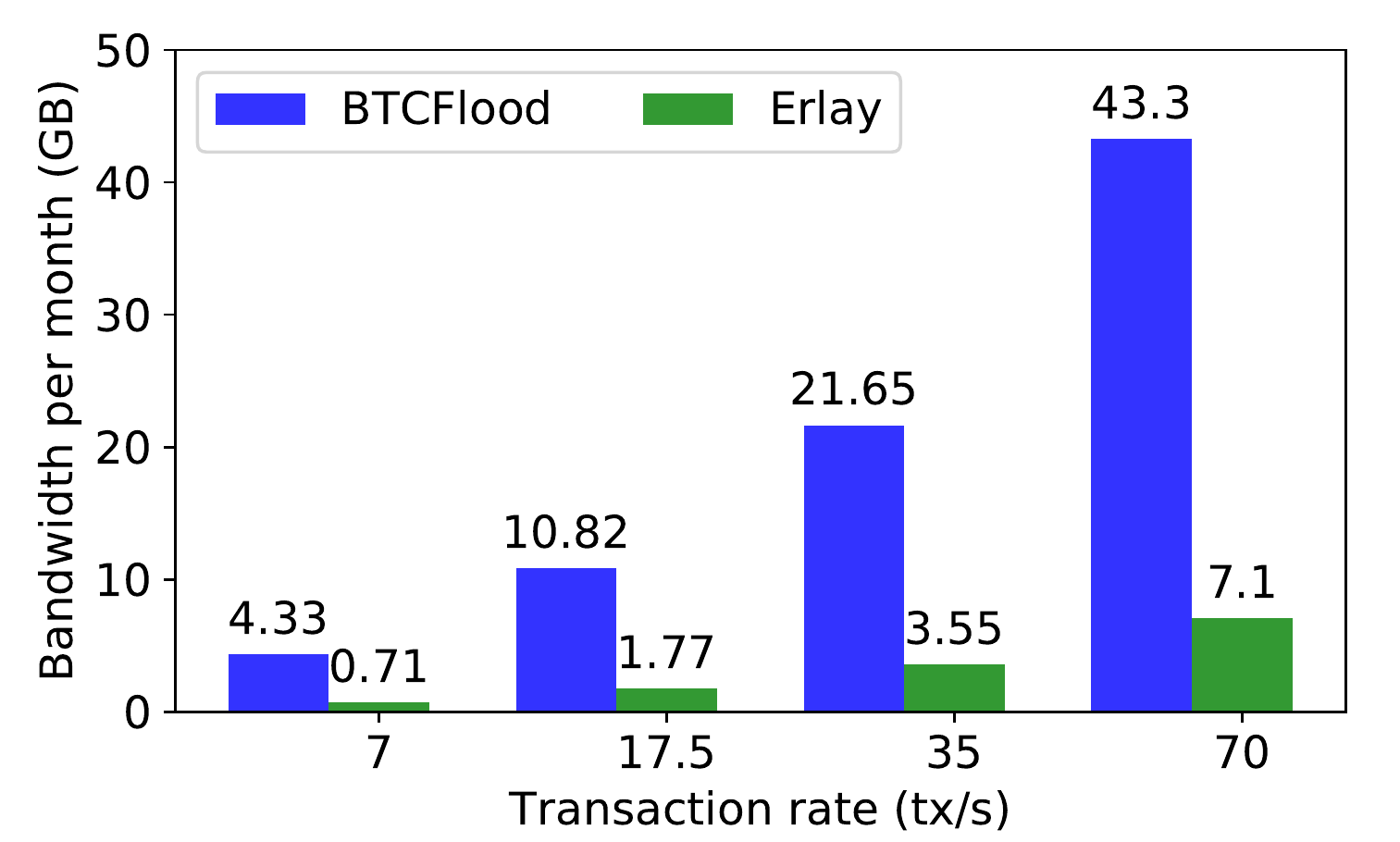}
\caption{Average bandwidth one node spends
  per month to \textbf{announce} transactions
  in a system with variable transaction rate}
\label{fig:bandwidth_txrate}
\end{figure}

\begin{figure}[t]
\centering
\includegraphics[width=\linewidth]{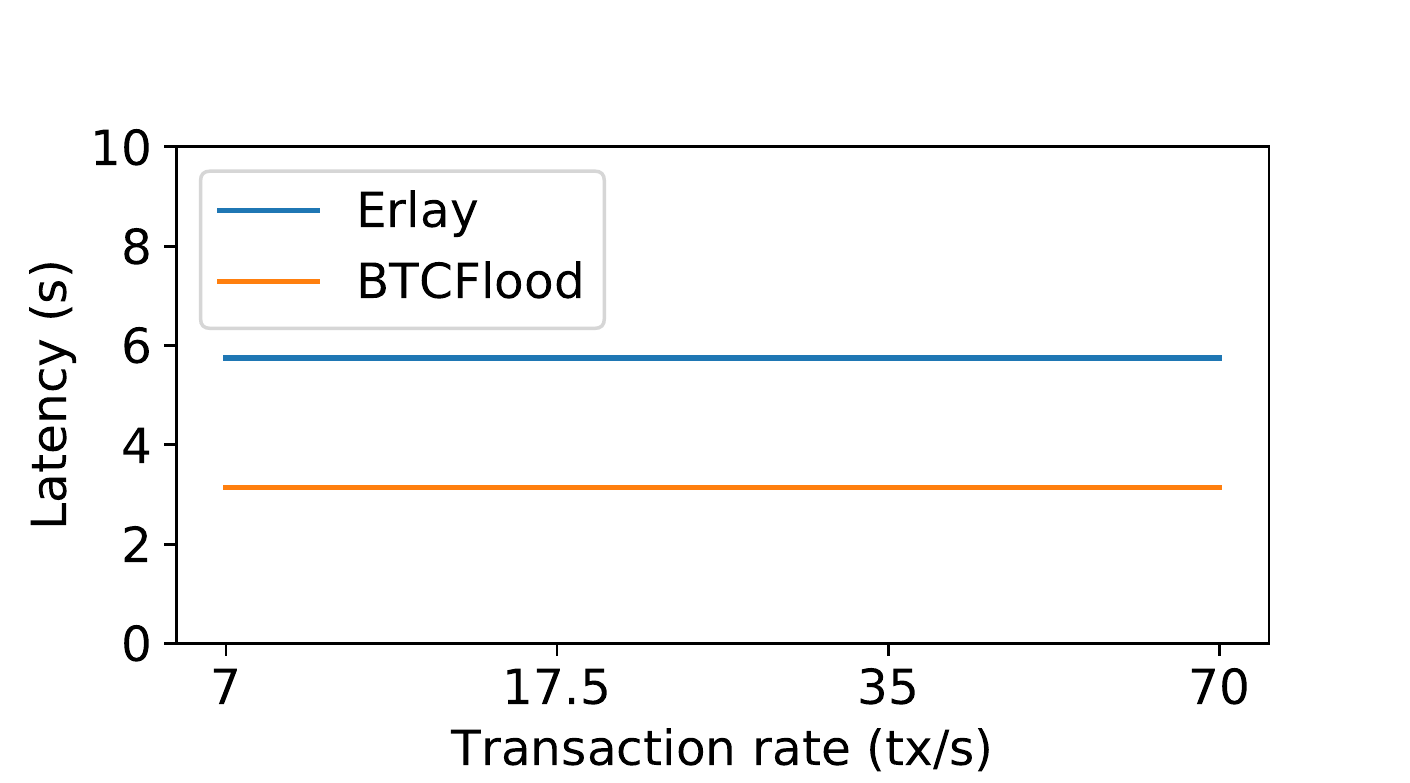}
\caption{Average latency for a single transaction to reach 100\% nodes
  in the network in a system with variable transaction rate}
\label{fig:latency_txrate}
\end{figure}

To demonstrate that bandwidth savings and latency are not impacted by
higher transaction rates, we simulated a network of 54,000 private and
6,000 public nodes with connectivity of 8, generated transactions at
different rates (from 7 tx/s to 70 tx/s), and measured the impact
of higher transaction rates on latency and bandwidth. 

Figure~\ref{fig:bandwidth_txrate} shows that the relative bandwidth
savings of Erlay is not impacted by transaction rate.
Figure~\ref{fig:latency_txrate} shows that Erlay's latency remains
constant for different transaction rates.  We also confirmed these results
in a network of 100 nodes running our prototype implementation.

\subsection{Withstanding timing attacks} \label{sec:withstand_timing}

One of Erlay's design goals is 
to be more robust to timing attacks from
sybils~\cite{Grundmann2018ExploitingTA, Segura2018Txprobe}.

To evaluate Erlay's robustness against timing attacks,
we simulated a network of 60,000 nodes and used \emph{first-spy estimator}
approach to link transactions to nodes of their origin.

With the first-spy estimator an attacker deploys some number of
\emph{spy} nodes. Each node keeps a local log of timestamped records,
each of which records (1) when the spy first learned about a
transaction, and (2) from which node the spy learned it. In our setup,
at the end of the experiment the spy nodes aggregate their logs and
estimate that the source node of a transaction is the node which was
the very first one to announce the transaction (to any of the spies).


Tables~\ref{table:private_spies} and~\ref{table:public_spies} list the
success rates of the first-spy estimator for different number of
spies, which were either private or public nodes.

While Erlay is more susceptible to spying by private nodes
(Tables~\ref{table:private_spies}), we believe that this is acceptable
for three reasons. (1) The success rate is below 50\% for both
protocols, which means that this deanonymization attack is unreliable,
(2) the difference between the two protocols is at most 10\%, and (3),
Erlay is materially more susceptible to spying when
there are higher levels of private spying nodes
(30\%). At this level, an attack with public spies is a more
reasonable alternative since the attacker must control fewer nodes to
achieve a higher attack success rate.

By contrast, Erlay increases the cost of the deanonymization attack by
public nodes (Table~\ref{table:public_spies}): an attacker must
control more long-running public nodes in the network with Erlay than
with BTCFlood to achieve the same attack rate.


\begin{table}[]
\begin{tabular}[c]{r||r|r}
Private node spies & \textbf{BTCFlood} & \textbf{Erlay} \\
\hline
5\%                      & 18\%     & 16\%  \\
10\%                     & 20\%     & 20\%  \\
30\%                     & 20\%     & 27\%  \\
60\%                     & 21\%    & 31\% 
\end{tabular}
\caption{Success rate of first-spy estimator with variable number
of private spying nodes in BTCFlood and Erlay.}
\label{table:private_spies}
\end{table}

\begin{table}[]
\begin{tabular}[c]{r||r|r}
Public node spies & \textbf{BTCFlood} & \textbf{Erlay} \\
\hline
5\%                      & 11\%     & 11\%  \\
10\%                     & 19\%     & 15\%  \\
30\%                     & 52\%     & 32\%  \\
60\%                     & 82\%    & 67\% 
\end{tabular}
\caption{Success rate of first-spy estimator with variable number
of public spying nodes in BTCFlood and Erlay.}
\label{table:public_spies}
\end{table}

We also measured that increasing the connectivity with Erlay
does not change success rate of first-spy estimation.

\subsection{Reconciliation and flooding trade-off}
\label{sec:tradeoff}

\begin{figure*}[t]
\centering
\includegraphics[width=\linewidth]{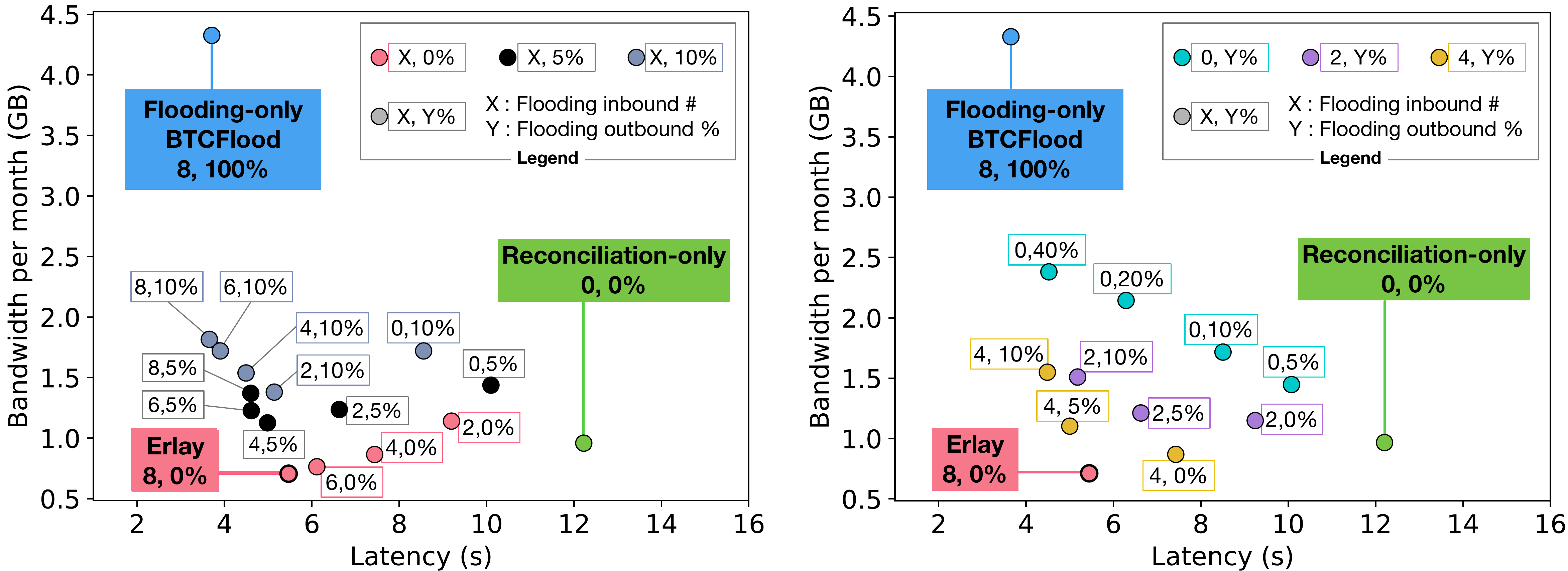}
\caption{Comparison of configurations of the Erlay-style protocol
  along the latency-bandwidth trade-off, as compared to BTCFlood
  (which does flooding only and no reconciliation). All points except
  for BTCFlood perform reconciliation on \emph{all} links. Each point
  varies the choice of the number of peers to \emph{flood} to that are
  inbound (out of 8 total), and outbound (out of 100\% total). Points
  with the same inbound/outbound configurations have the same
  color. We split the points across two plots for readability.}
\label{fig:config_perf}
\end{figure*}

Erlay's design combines flooding with reconciliation to achieve a
balance between two extremes: the current flooding-only protocol in
Bitcoin (BTCFlood), and a reconciliation-only protocol. This intuition
is captured in the latency-bandwidth trade-off diagram in
Figure~\ref{fig:bw_latency_tradeoff}. However, does Erlay actually
strike a balance? And, what other intermediate protocol alternatives
lie between flooding-only and reconciliation-only designs?

A key design choice in Erlay is to flood transactions to 8 outbound
peers and none to the inbound peers. We have also considered other
alternatives while designing Erlay. Although a full exposition of the
design space is beyond the scope of this paper, we present a limited
comparison of the latency-bandwidth trade-off for several other
protocol variants that use a different choice of flooding
inbound/outbound peers. Specifically, we used our simulator to collect
data about versions of the Erlay protocol that use $X$ inbound peers
and $Y$ outbound peers for flooding (while using reconciliation on all
links including $X$ and $Y$), for different values of $X$ and $Y$.


We ran several experiments, with each experiment being a protocol
configuration that select a specific $X$ inbound and $Y$ outbound
values. In these experiments we simulated a network of 24,000 private
and 6,000 public nodes and relayed a total of 1,000
transactions\footnote{We restricted the network size to constraint the
  experiment running time}. We collected transaction latency and
bandwidth usage for each experiment and Figure~\ref{fig:config_perf}
plots the results.

Figure~\ref{fig:config_perf} shows that BTCFlood and
Reconciliation-only indeed lie at opposite ends of the trade-off
spectrum (top left for BTCFlood and bottom right for
Reconciliation-only). And, most key, Erlay lies closer to the bottom
left corner than either configuration. This figure also shows that
configurations with other choices of values for $X$ and $Y$ get close
to the left corner. But they do not strike as good a balance between
latency and bandwidth as Erlay does.


\section{Reference implementation results}
\label{sec:ref_impl_res}

We implemented Erlay as part of Bitcoin Core. For this we added 584 LOC,
not including Minisketch.
We used a network of 100 Azure nodes located in 6 data centers,
running a reference implementation of our protocol integrated in Bitcoin Core node
software, to evaluate Erlay in deployment.
We generated and relayed
1000 transactions, all originating from one node with a rate of 7
transactions per second.
We compared the average latency and bandwidth of Erlay versus
Bitcoin's current implementation. Table~\ref{table:impl_results}
summarizes our results.
According to our measurements, Erlay introduced a latency increase of
0.2 seconds, while saving 40\% of the overall node bandwidth.

As in our simulations, Erlay has a higher latency but lower
bandwidth cost, confirming our original design intent
(Fig.~\ref{fig:bw_latency_tradeoff}).

\begin{table}[t]
\begin{tabular}{r||r|r}
                                                                 & \textbf{BTCFlood} & \textbf{Erlay}   \\
\hline
\begin{tabular}[c]{@{}l@{}}Base cost (MB)\\ (TX+GETDATA)\end{tabular} & 27  & 27 \\
Other messages (MB)                                           & 1.06  & 1.1 \\
Announcement cost (MB)                                           & 42  & 15 \\
Latency (s)                                               & 1.85      & 2.05
\end{tabular}
\caption{Prototype measurements collected from a 100-node deployment
  comparing the latency and bandwidth of the BTCFlood in the reference
  implementation against our Erlay implementation.
  \label{table:impl_results}
}
\end{table}

\section{Discussion}
\label{sec:discussion}

\textbf{Reconciliation-only relay.}
We believe that a reconciliation-only transaction relay protocol would
be inherently susceptible to timing attacks that could reveal the
source of the transaction.
Unlike flooding, reconciliation is inherently bi-directional:
an inbound connection for one peer is an outbound connection for
another peer. Delays cannot be applied per-direction but
rather per-link. Therefore, BTCFlood's diffusion delay cannot be used
in reconciliation.

\textbf{Erlay increases latency from 3.15s to 5.75s}
Erlay increases the \emph{time to relay an unconfirmed transaction across all nodes},
which is a small fraction of the end-to-end transaction processing (10 minutes).

We tuned Erlay to maximize bandwidth savings assuming that
an increase in latency from 3.15s to 5.75s is acceptable.
It is possible to tune Erlay to provide the same latency as
BTCFlood by reconciling more often, but this would save 70\% of transaction relay
bandwidth instead of 84\%. If we tuned Erlay to provide the same latency as
BTCFlood, we could increase network connectivity and improve the network
security without additional bandwidth overhead.

In practice, there are 2 primary implications of 
\emph{transaction relay latency} increase.

\emph{Block production rate} is defined by block relay latency, which is only
indirectly defined by transaction relay latency: if fewer transactions are relayed,
it will take longer for blocks to propagate
(since missing transactions have to be relayed and validated).
Block production rate is defined in this way
because to maximize the security of the network all miners
have to work on the latest block and can avoid generating ``orphan'' blocks.
Because Erlay's latency among public nodes is better than
BTCFlood (Erlay's diffusion interval is lower), miners' orphan rate
will probably be lower with Erlay. And, because most miners today use
an overlay network (e.g., FIBRE), transaction relay latency increase (3.15s to 5.75s
with Erlay will have even less impact.

\emph{User experience.} If a transaction is accepted in an unconfirmed state,
then the user perceives the 2.6s latency increase.
However, unconfirmed transactions are rarely accepted by users.
Instead, users wait for at least 10 minutes to confirm transactions.
Therefore, we think that Erlay's 2.6s latency increase insignificantly
impacts the users' experience.

\textbf{Compatibility with Dandelion.}
Dandelion is an alternative transaction relay protocol introduced to
improve the anonymity and robustness to adversarial observers in
Bitcoin~\cite{Fanti2018Dandelion++}. Dandelion has two phases: stem
(propagation across a single link of ten nodes on average), and fluff
(relay using flooding from the last node in the stem link).
Erlay is complimentary with Dandelion: Erlay would replace the fluff phase in
Dandelion, while the stem phase of Dandelion would flood through both
inbound and outbound links to preserve the privacy of private nodes.

\textbf{Backward compatibility.}
Only about 30\% of Bitcoin nodes run the latest release
of Bitcoin Core\footnote{\url{https://luke.dashjr.org/programs/bitcoin/files/charts/security.html}}.
Therefore, Erlay must be backwards compatible. 
%
%
If not all nodes use Erlay, then Erlay may be activated
per-link if both peers support it.






%



\textbf{Sophisticated timing attacks.}
In Section \ref{sec:withstand_timing} we demonstrated that Erlay
is less susceptible to timing attacks based on the first-spy estimator. 
Withstanding more sophisticated attacks (e.g., fingerprinting
propagation traces) is an open question for future research.





\textbf{Mining-related attacks.}
There is no direct relationship between Erlay and attacks
like selfish mining~\cite{Eyal2013Majority}. By making timing attacks more expensive,
Erlay makes it harder to infer the network topology. Inferring
the topology would allow clustering the network by attacking bottlenecks.
Clustering the network would then split mining efforts and
introduce many orphan blocks until the network clusters recompose.
Thus, Erlay indirectly makes the network stronger.

\textbf{Relevance to other blockchains.}  Erlay is relevant to most
other deployed blockchains (e.g., Ethereum, Zcash) because they use
flooding for transaction relay. Even though there might be a
difference in TXID size or number of connected peers, the difference
that matters is transaction rate. As
Figures~\ref{fig:bandwidth_txrate} and~\ref{fig:latency_txrate}
illustrate, Erlay is theoretically suitable for systems with higher
transaction rate.

On the other hand, since PinSketch has quadratic complexity, using it
without modifications would lead to the high computational cost of reconciliation,
and higher hardware requirements. To reduce the computational cost of
reconciliation, we suggest applying bisection from the first reconciliation step.

For example, consider a system with a network similar to Bitcoin,
but with a throughput of 700 transactions/s. If Erlay is
applied in the same way as we suggest for Bitcoin, an average
reconciliation set difference would consist of 1,000
elements. According to the benchmarks, straightforward reconciliation
through Minisketch would take 1,000 ms. At the same time, with
bisection recursively applied 3 times, 8 chunks consisting of 125
elements would have to be reconciled, and this would take only 20
ms. This result makes Erlay usable for systems with much higher
transaction rate.

We do not propose this measure for Bitcoin, because considering the transaction rate
in Bitcoin, the computational cost of reconciliation is already low enough.

\section{Related Work}
\label{sec:related}

Prior studies of Bitcoin's transaction relay focused on information
leakage and other vulnerabilities~\cite{Fanti2018Dandelion++,
  Neudecker2016BTCTimingAnalysis}, and did not consider bandwidth
optimization. We believe that our work is the first to introduce a
bandwidth-efficient, low-latency, and robust transaction relay
alternative for Bitcoin.
Erlay is designed as a minimal change to Bitcoin (584 LOC),
in contrast with other proposals that optimize Bitcoin more
deeply~\cite{Eyal2016BitcoinNG}.

%

\noindent \textbf{Short transaction identifiers.}
One solution to BTCFlood's inefficiency is to use \emph{short
  transaction identifiers}.
There are two issues with this solution. First, this
\emph{only reduces bandwidth cost by a constant factor}. In our
simulation we found that short identifiers would reduce redundant
traffic from 43\% to 10\%. But, with higher connectivity, redundancy
climbs back up faster than it does with Erlay. The second issue with
short IDs is that they would make the system vulnerable to
collision-related attacks, requiring a new per-node or per-link
secure salting strategy.

\noindent \textbf{Blocksonly setting.}
Bitcoin Core 0.12 introduced a \emph{blocksonly} setting in which a
node does not send or receive individual transactions; instead, the
node only handles complete blocks. As a result, blocksonly has no INV
message overhead.
%
%
In the blocksonly case, nodes will have to relay and receive many
transactions at once. This will increase the maximum node bandwidth
requirements and cause spikes in block content relay and
transaction validation. 
\noindent \textbf{Reconciliation alternatives.}
Prior work has also devised multi-party set
reconciliation~\cite{Mitzenmacher2013MultiPartyRecon,
  Mitzenmacher2014MultiPartyRecon}. This approach, however, has
additional complexity and additional trust requirements between
peers. We believe that the benefits of such an approach are not
substantial enough to justify these limitations.

In addition, reconciliation-based techniques usually provide
bandwidth-efficiency under the assumptions where most of the state
being reconciled is shared~\cite{Corallo2016CompactBlocks,
  Ozisik2017Graphene}.

\textbf{Network attacks on Bitcoin and connectivity.} The security of
the Bitcoin network has been under substantial scrutiny with many
published network-related attacks~\cite{Bonneau2016WhyBW,
  McCorry2016RefundAO, narayanan2016bitcoin, Miller2013FeatherFork,
  Andrychowicz2015Malleability, Decker2014Malleability,
  corbixgwelt2011timejacking, Johnson2014GameTheorMining,
  Douceur2002Sybil, Heilman2015Eclipse, Gervais2015TamperingBlocks,
  Apostolaki2017Hijack, Koshy2014BTCAnonymity, Biryukov2014DeanonBTC}.
These attacks attempt to make the network weaker (e.g., increase the
probability of double-spending or denials of service) or violate user
privacy. Many of these attacks rely on non-mining nodes and assume
limited connectivity from victim nodes. Our work allows Bitcoin nodes
to have higher connectivity, which we believe will make the
network more secure.

\noindent \textbf{Prior P2P research.}
Structured P2P networks are usually based on Distributed Hash Tables
(DHTs), in which every peer is responsible for specific
content~\cite{Maymounkov2002Kademlia}. In these networks research has
explored the use of topology information to make efficient routing
decisions~\cite{Stoica2001Chord, Rowstron2001Pastry,
  Vuong2003EffRouting, Clarke2001Freenet}. This design, however, makes
these protocols leak information about the structure of the network
and makes them less robust to Byzantine faults, though \emph{limited}
solutions to Byzantine faults in this setting have been
explored~\cite{Fiat2005ChordRobust, Dearle2010ByzantineChordRing}.

The trade-off between latency and bandwidth efficiency is well-known
in P2P research. Kumar et. al. identified and formalized the trade-off
between latency and bandwidth~\cite{Kumar2005BWLatencyDHT}, and Jiang
et. al. proposed a solution to achieve an optimal combination of these
properties~\cite{Jiang2010BWLatencyFriendcast}. However, the solution
was not designed for adversarial settings.

Prior work also proposed feedback-based approaches to
flooding~\cite{Papadakis2007FeedbackFlood, AI2006EffFlood}. However,
we believe that to work efficiently (have a \emph{horizon} larger than
1), this work would have unacceptable information leakage.




\section{Conclusions}
\label{sec:conclusion}

Bitcoin is one of the most widely used P2P applications.
Today, Bitcoin relies on flooding to relay transactions in a network
of about 60,000 nodes. Flooding provides low latency and is robust to
adversarial behavior, but it is also bandwidth-inefficient and creates
a significant amount of redundant traffic.
We proposed Erlay, an alternative protocol that combines limited
flooding with intermittent reconciliation. We evaluated Erlay in
simulation and with a practical deployment.
Compared to Bitcoin's current protocols, Erlay reduces the bandwidth
used to announce transactions by 84\% while increasing the latency for
transaction dissemination by 2.6s (from 3.15s to 5.75s).
Erlay allows Bitcoin nodes to have higher connectivity, which will
make the network more secure. We are actively working to introduce
Erlay into Bitcoin Core's node software.


\section*{Acknowledgments}
\label{sec:acks}

Gleb Naumenko was partially supported by Blockstream. This research
was also funded by an NSERC discovery grant to Ivan Beschastnikh and
was supported in part by a gift from Microsoft Azure.

\bibliographystyle{ACM-Reference-Format}
\bibliography{paper}

\end{document}